\documentclass[traditabstract]{aa}
\usepackage{graphicx}
\usepackage{txfonts}
\usepackage{color}

\newcommand{\obj}{H1413+117}

\begin{document}

\title{Microlensing in H1413+117 : disentangling line profile emission
and absorption in a broad absorption line quasar\thanks{Based on
observations made with the Canada-France-Hawaii Telescope (Hawaii),
with ESO Telescopes at the Paranal Observatory (Chile) and with the
NASA/ESA Hubble Space Telescope, and obtained from the data archive at
the Space Telescope Institute. ESO program ID: 074.A-0152, 075.B-0675,
081.A-0023.}}

\author{D. Hutsem\'ekers\inst{1,}\thanks{Ma{\^\i}tre de Recherches du F.N.R.S.}
   \and B. Borguet\inst{1,}\thanks{Boursier du F.N.R.S.} 
   \and D. Sluse\inst{2} 
   \and P. Riaud\inst{1} 
   \and T. Anguita\inst{2,3}}

\institute{Institut d'Astrophysique et de G\'eophysique,
           Universit\'e de Li\`ege, All\'ee du 6 Ao\^ut 17, B5c, B-4000
           Li\`ege, Belgium
      \and Astronomisches Rechen-Institut, Zentrum f\"{u}r Astronomie der 
           Universit\"{a}t Heidelberg (ZAH),  M\"{o}nchhofstr.\ 12-14, 
           69120 Heidelberg, Germany
      \and Departamento de Astronom\'{i}a y Astrof\'{i}sica, Pontificia 
           Universidad Cat\'{o}lica de Chile, Santiago, Chile}
\date{Received ; accepted: }

\titlerunning{Microlensing in the BAL QSO H1413+117} 
   

\abstract{
On the basis of 16 years of spectroscopic observations of the four
components of the gravitationally lensed broad absorption line (BAL)
quasar \obj , covering the ultraviolet to visible rest-frame spectral
range, we analyze the spectral differences observed in the
P~Cygni-type line profiles and have used the microlensing effect to
derive new clues to the BAL profile formation. We first find that the
absorption gradually decreases with time in all components and that
this intrinsic variation is accompanied by a decrease in the intensity
of the emission. We confirm that the spectral differences observed in
component D can be attributed to a microlensing effect lasting at
least a decade.  We show that microlensing magnifies the continuum
source in image D, leaving the emission line region essentially
unaffected. We interpret the differences seen in the absorption
profiles of component D as the result of an emission line superimposed
onto a nearly black absorption profile.  We also find that the
continuum source and a part of the broad emission line region are
likely de-magnified in component C, while components A and B are not
affected by microlensing. Differential dust extinction is measured
between the A and B lines of sight.  We show that microlensing of the
continuum source in component D has a chromatic dependence compatible
with the thermal continuum emission of a standard Shakura-Sunyaev
accretion disk.  Using a simple decomposition method to separate the
part of the line profiles affected by microlensing and coming from a
compact region from the part unaffected by this effect and coming from
a larger region, we disentangle the true absorption line profiles from
the true emission line profiles. The extracted emission line profiles
appear double-peaked, suggesting that the emission is occulted by a
strong absorber, narrower in velocity than the full absorption
profile, and emitting little by itself.  We propose that the outflow
around \obj\ is constituted by a high-velocity polar flow and a
denser, lower velocity disk seen nearly edge-on. Finally, we report on
the first ground-based polarimetric measurements of the four
components of \obj .}

   
\maketitle

%

\section{Introduction}
\label{sec:intro}

The broad absorption lines (BALs) observed in the spectra of quasars
(or QSOs, quasi-stellar objects), blueshifted with respect to the
broad emission lines (BELs), reveal massive, high-velocity outflows in
active galactic nuclei (AGN). Such powerful winds can strongly affect
the formation and evolution of the host galaxy, enrich the
intergalactic medium, and regulate the formation of the large-scale
structures (e.g.  Silk and Rees \cite{sil98}, Furlanetto and Loeb
\cite{fur01}, Scannapieco and Oh \cite{sca04}, Scannapieco et
al. \cite{sca05}).

About 15\% of optically selected quasars have BALs in their spectra
(Reichard et al. \cite{rei03}). Outflows may be present in all quasars
if the wind is confined into a small solid angle so that BALs are only
observed when the flow appears along the line of sight (Weymann et
al. \cite{wey91}, Gallagher et al. \cite{gal07}). On the other hand,
BAL QSOs could be quasars in an early evolutionary stage, washing out
their cocoons (Voit et al. \cite{voi93}, Becker et al. \cite{bec00}).

Despite many high-quality observational studies, in particular from
spectropolarimetry (Ogle et al. \cite{ogl99}), no clear view of the
geometry and kinematics of the BAL phenomenon has emerged yet.  While
pure spherically symmetric winds appeared too simple to account for
the variety of observations (Hamann et al. \cite{ham93}, Ogle et
al. \cite{ogl99}), equatorial disks, rotating winds, polar flows, or
combinations thereof have been proposed more or less successfully to
interpret the observations of individual objects or small groups of
them (e.g. Murray et al. \cite{mur95}, Schmidt and Hines \cite{sch99},
Lamy and Hutsem\'ekers \cite{lam04}, Zhou et al. \cite{zho06}).  Given
the large parameter space characterizing non spherically symmetric
winds, BAL profile modeling must then be combined with other
techniques to determine the outflow properties in individual objects
(e.g.  Young et al. \cite{you07}).

An interesting method that can bring independent information on the
quasars' internal regions is the use of gravitational microlensing.
Indeed, in a typical gravitationally lensed quasar, a solar mass star
belonging to the lensing galaxy has an Einstein radius $R_E$ (the
microlensing cross section) on the order of 10$^{-2}$~pc, which is
comparable to the size of the continuum source. The microlens, moving
across the quasar core in projection, can successively magnify regions
of area $\simeq$ $\pi R^2_E$, inducing spectroscopic variations that
could be used to extract information on the quasar structure
(Schneider et al. \cite{sch92}, and references therein). Several
studies, based on simulations, have demonstrated the interest of
microlensing analyses for understanding BAL QSOs (Hutsem\'ekers et
al. \cite{hut94}, Lewis and Belle \cite{lew98}, Belle and Lewis
\cite{bel00}, Chelouche \cite{che05}).

\begin{figure}[t]
\resizebox{\hsize}{!}{\includegraphics*{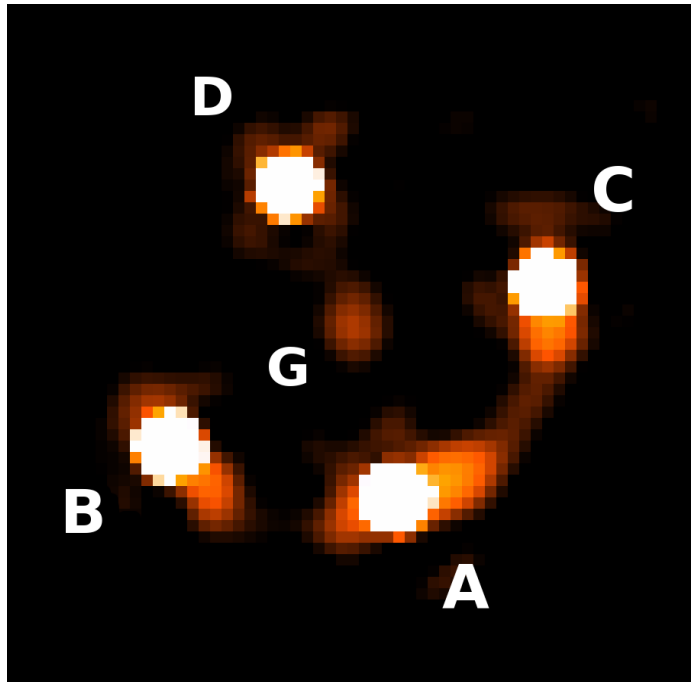}}
\caption{A deconvolved near-infrared image of the gravitationally
lensed quasar \obj\ with the four images and the lensing galaxy
labelled (from Chantry and Magain \cite{cha07}). The image has been
obtained in the F160W filter ($\lambda \simeq$ 1.6 $\mu$m) with the
NICMOS camera attached to the Hubble Space Telescope. North is up and
East to the left. The angular separation between components A and D is
1$\farcs$1.}
\label{fig:image}
\end{figure}

\obj\ is a BAL QSO of redshift $z \simeq 2.55$ showing typical
P~Cygni-type profiles, i.e., profiles where the absorption is not
detached from the emission. Turnshek et al. (\cite{tur88}) discussed
the spectrum of \obj\ in detail and made the first attempts to
disentangle the emission from the absorption assuming an intrinsic
blue/red symmetry of the emission lines.  \obj\ is also a
gravitationally lensed quasar constituted of four images (Magain et
al. \cite{mag88}; see Fig.~\ref{fig:image}).  The lensing galaxy is
faint and its redshift poorly known: indirect estimates give $z_l
\simeq 1.0$ (Kneib et al. \cite{kne98}) or $z_l \simeq 1.88$
(Goicoecha and Shalyapin \cite{goi10}).  Evidence of microlensing in
component D has been suggested from both photometry and spectroscopy
(Angonin et al. \cite{ang90}, {\O}stensen et al. \cite{ost97}). In
particular, Angonin et al. (\cite{ang90}) found that the equivalent
width of the emission lines is systematically smaller in component D
than observed in the other components, a result which can be
interpreted by microlensing of the continuum source, the larger region
at the origin of the emission lines being unaffected. This effect,
which appeared to last at least a decade (Chae et al. \cite{cha01},
Anguita et al. \cite{ang08}), offers the possibility to separate the
microlensed attenuated continuum (i.e. the absorption profile) from
the true emission line profile, thus providing new clues to the
formation of BAL profiles (Hutsem\'ekers \cite{hut93}, Hutsem\'ekers
et al. \cite{hut94}).

In the present paper, we homogeneously analyze the spectra of the four
components of \obj\ obtained from 1989 to 2005. The spectra cover the
ultraviolet to visible rest-frame spectral range.  In
Sect.~\ref{sec:spectro}, we show that the spectral differences
observed between the images can be consistently attributed to
microlensing in spite of intrinsic variations. In
Sect.~\ref{sec:decomp}, using a simple method, we separate the parts
of the spectra affected and unaffected by microlensing, which
basically correspond to the attenuated continuum and the emission
lines. From these results, we derive a consistent view of the macro-
and microlensing in \obj\ (Sect.~\ref{sec:lensing}). Finally, with the
``pure'' absorption and emission profiles in hand, we discuss the
formation of the BAL profiles and the implications for the geometry
and the kinematics of the outflow (Sect.~\ref{sec:bal}).

\section{Data collection}
\label{sec:data}

Spectra of the four components of \obj\ were gathered from archived
and published data. Table~\ref{tab:spectro} summarizes the
characteritics of the spectra obtained over a period of 16 years, with
the date of observation, the spectral range, the average resolving
power $R = \lambda / \Delta \lambda$, and the instrument used.

The visible spectra secured in 1989 with the bidimensional
spectrograph SILFID at the Canada-France-Hawaii Telescope (CFHT) are
described in Angonin et al.  (\cite{ang90}) and Hutsem\'ekers
(\cite{hut93}). These spectra were obtained under optimal seeing
conditions (0$\farcs$6 FWHM).  They provided the first spectroscopic
evidence of microlensing in \obj .

A series of spectra were obtained in 1993-1994 with the Hubble Space
Telescope (HST) feeding the Faint Object spectrograph (FOS). They
cover the UV-visible spectral range (gratings G400H and G570H). These
data are described in Monier et al.  (\cite{mon98}).  A second series
of HST spectra, yet unpublished, were obtained in 2000 using the Space
Telescope Imaging Spectrograph (STIS) and the G430L grating (principal
investigator : E. Monier; proposal \# 8127). All HST data were
retrieved from the archive and reduced using standard procedures for
long slit spectroscopy and prescriptions by Monier et
al. (\cite{mon98}).

In 2005, visible spectra were obtained with the integral field unit of
the Visible MultiObject Spectrograph (VIMOS) attached to the European
Southern Observatory (ESO) Very Large Telescope (VLT). The data were
obtained under medium quality seeing conditions (1$\farcs$2
FWHM). Details on the observations and reductions are given in Anguita
et al. (\cite{ang08}). For this data set, it was not possible to
separate the spectra of images A and B of \obj .

The 2005 near-infrared spectra obtained with the integral field
spectrograph SINFONI at the VLT were retrieved from the ESO archive
(principal investigator : A. Verma ; proposal 075.B-0675(A)). Only the
spectra obtained with the best seeing (0$\farcs$5 FWHM on May 22, 2005
in the H and K spectral bands) are considered here. The pixel size was
0$\farcs$125$\times$0$\farcs$250 on the sky. The observations consist
of four exposures per spectral band. The object was positionned at
different locations on the detector for sky subtraction.  The data
were reduced using the SINFONI pipeline.  Telluric absorptions were
corrected using standard star spectra normalized to a blackbody. The
individual spectra were extracted by fitting a 4-gaussian function
with fixed relative positions and identical widths to each image plane
of the data cube using a modified MPFIT package (Markwardt
\cite{mar09}). Astrometric positions were taken from HST observations
(Chantry and Magain \cite{cha07}). The spectra which appeared affected
by detector defects and/or important cosmic ray hits were
discarded. The good spectra were finally filtered to remove remaining
spikes.

In addition, we have observed \obj\ on May 10, 2008 with the
polarimetric mode of the Focal Reducer and low dispersion Spectrograph
1 (FORS1) installed at the Cassegrain focus of the VLT. Observations
have been carried out with the V$_{\rm high}$ filter, under excellent
seeing conditions (0$\farcs$6). Linear polarimetry has been performed
by inserting in the parallel beam a Wollaston prism, which splits the
incoming light rays into two orthogonally polarized beams, and a
half-wave plate rotated to four position angles (e.g. Sluse et
al. \cite{slu05}). In order to measure the polarization of the four
images, the MCS deconvolution procedure devised by Magain et
al. (\cite{mag98}) has been applied. We used a version of the
algorithm that allows for a simultaneous fit of different individual
frames obtained with the same observational setup (Burud
\cite{bur01}). We constructed the PSF using a bright point-like object
located $\sim$ 15\arcsec\ from \obj . Since this object is close to
our target and is similar in brightness to the individual components
of \obj , it provided a good estimate of the PSF. The Stokes
parameters have been calculated from the photometry of the quasar
lensed images derived from the deconvolution process.

\begin{table}[t]
\caption{Spectroscopic data}
\label{tab:spectro}
\begin{tabular}{lccl}\hline\hline \\[-0.10in]
   Date &  Spectral range &  $R$  & Instrument \\ 
   (y/m/d) &    &   &   \\ 
\hline \\[-0.10in]
 1989/03/07    & 4400--6700 \AA    &   $\sim$ 450     & CFHT + SILFID \\ 
 1993/06/23    & 4600--6800 \AA    &   1300           & HST + FOS  \\ 
 1994/12/24    & 3250--4800 \AA    &   1300           & HST + FOS  \\ 
 2000/04/21-26 & 3000--5700 \AA    &   500--1000      & HST + STIS  \\ 
 2005/03/18    & 3700--6700 \AA    &   $\sim$ 250     & VLT + VIMOS  \\ 
 2005/05/22    & 1.95--2.45 $\mu$m &   $\sim$ 4000    & VLT + SINFONI \\
 2005/05/22    & 1.45--1.85 $\mu$m &   $\sim$ 3000    & VLT + SINFONI \\
 2005/06/07    & 1.10--1.40 $\mu$m &   $\sim$ 2000    & VLT + SINFONI \\
\hline\\[-0.2cm]
\end{tabular}
\end{table}

\begin{figure*}[!]
\resizebox{16.5cm}{!}{\includegraphics*[angle=90]{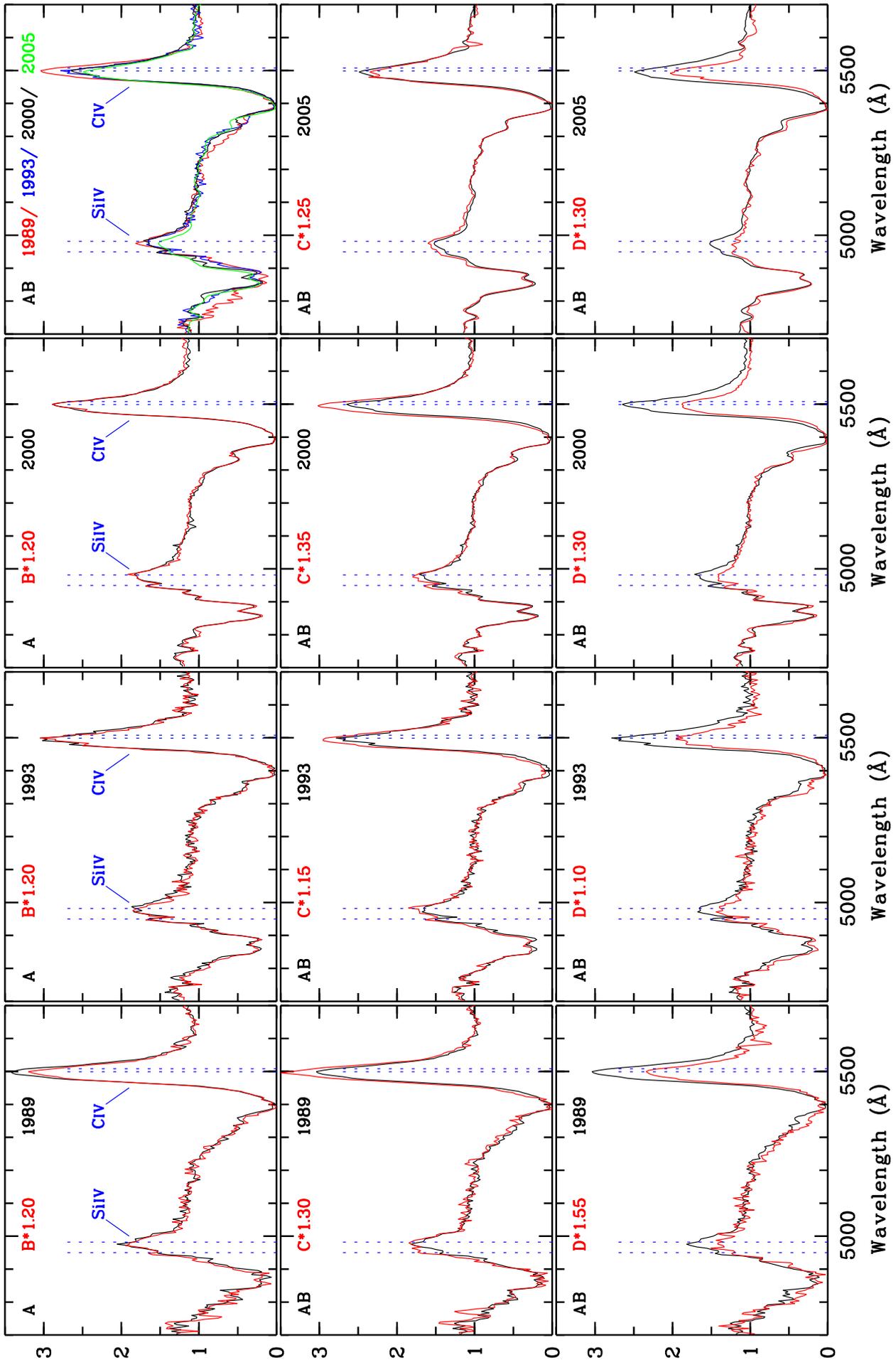}}
\caption{Intercomparison, at different epochs, of the \ion{Si}{iv} and
\ion{C}{iv} line profiles illustrating the spectral differences
between some images of \obj . Ordinates are relative fluxes. Vertical
dotted lines indicate the positions of the spectral lines at the
redshift $z = 2.553$ . The scaling factors needed to superimpose the
continua are indicated.  AB refers to the average spectrum of the A
and B components (which could not be separated in the 2005 spectra).
The upper left panel illustrates the time variation of the AB
spectrum.}
\label{fig:specvis}
\end{figure*}

\begin{figure*}[!]
\resizebox{16.0cm}{!}{\includegraphics*[angle=90]{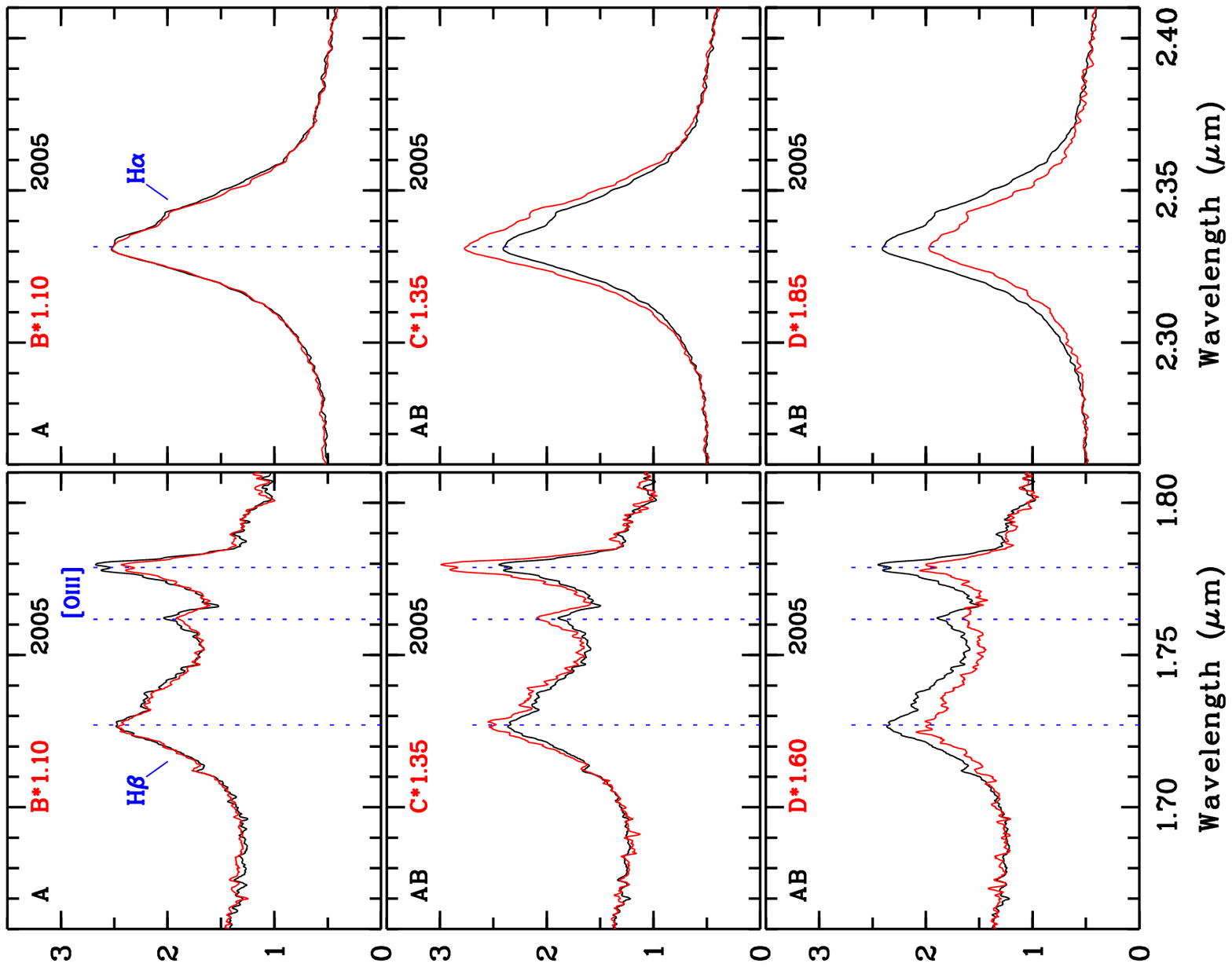}}
\resizebox{16.0cm}{!}{\includegraphics*[angle=90]{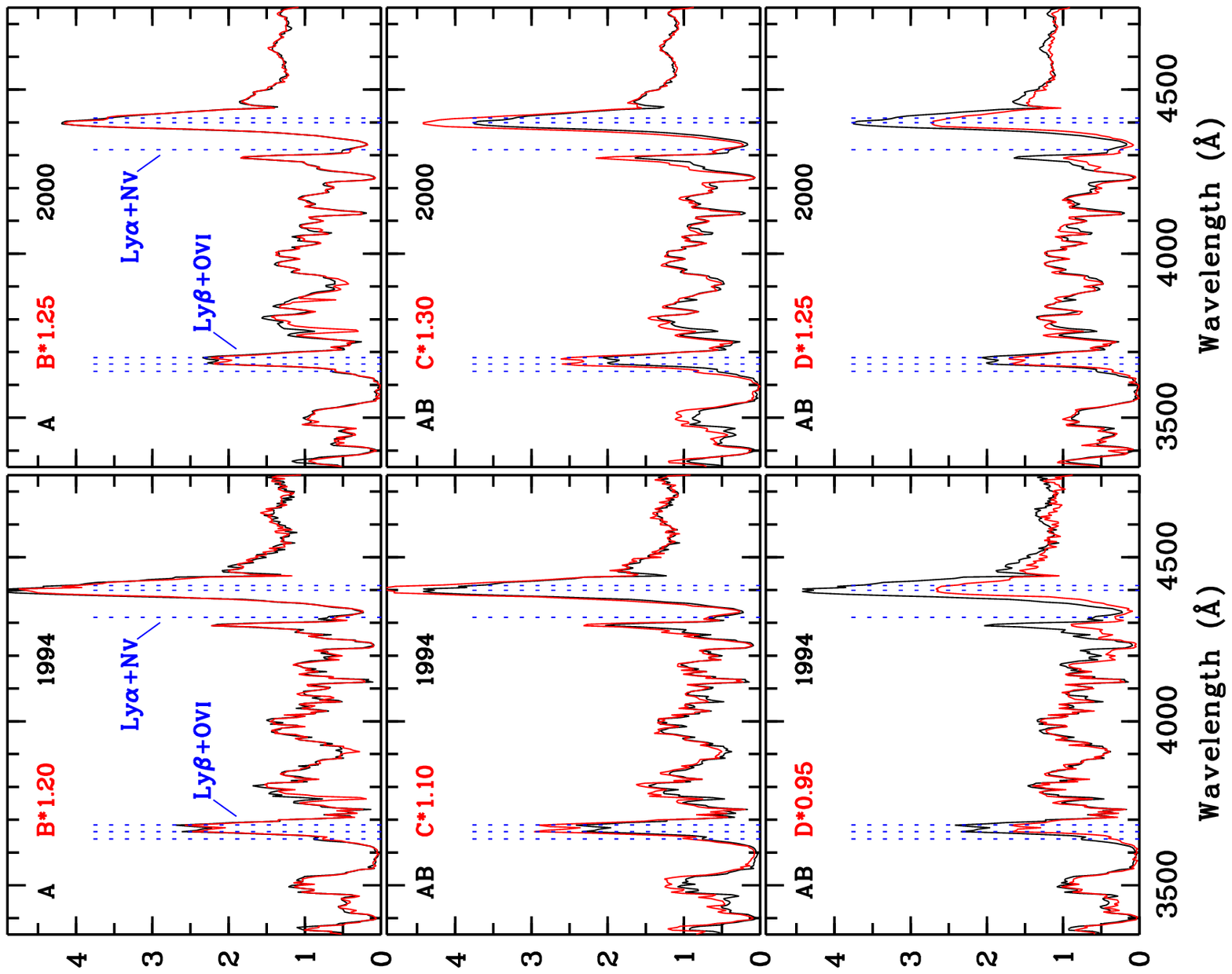}}
\caption{Intercomparison of the Ly$\beta$ + \ion{O}{vi} and Ly$\alpha$
+ \ion{N}{v} line profiles in the 4 images of \obj\ at two epochs
(lower panels). The comparison of the H$\beta$ + [\ion{O}{iii}] and
H$\alpha$ line profiles is shown in the upper panels. As in
Fig.~\ref{fig:specvis}, ordinates are relative fluxes and vertical
dotted lines indicate the redshifted line positions.  The scaling
factors needed to superimpose the continua are indicated. }
\label{fig:specuvir}
\end{figure*}

\section{Description of the spectra}
\label{sec:spectro}

At the redshift of \obj , spectra obtained in the UV-visible contain
\ion{C}{iv} $\lambda\lambda$1548,1550, \ion{Si}{iv}
$\lambda\lambda$1393,1402, \ion{N}{v} $\lambda\lambda$1238,1242,
\ion{P}{v} $\lambda\lambda$1117,1128, \ion{O}{vi}
$\lambda\lambda$1031,1037, Ly$\alpha$ $\lambda$1216 and Ly$\beta$
$\lambda$1026, while the near-infrared spectra contain H$\alpha$
$\lambda$6563, H$\beta$ $\lambda$4861 and [\ion{O}{iii}]
$\lambda\lambda$4959,5007 (Figs. \ref{fig:specvis} and
\ref{fig:specuvir}). From the [\ion{O}{iii}] lines we measure the
redshift $z = 2.553$.  The UV resonance lines show typical P
Cygni-type profiles with deep absorption while the Balmer lines show
broad emission possibly topped with a narrower feature.  Ly$\alpha$
and Ly$\beta$ lines are weak due to absorption by the \ion{N}{v} and
\ion{O}{vi} ions, respectively (e.g. Surdej and Hutsem\'ekers
\cite{sur87}).

\subsection{Evidence for microlensing}
\label{sec:microe}

In Figs.~\ref{fig:specvis} and \ref{fig:specuvir}, we compare the
profiles of various spectral lines observed in the different images A,
B, C, and D of \obj . A scaling factor is applied to superimpose at
best the continua, considering in particular the continuum windows
4525--4545 \AA\ and 5165--5220 \AA\ (i.e. 1275--1280 \AA\ and
1450--1470 \AA\ rest-frame, Kuraszkiewicz et al. \cite{kur02}).  If
the four images are only macrolensed, the line profiles observed in
the spectra of the four components must be identical up to the scaling
factor.  On the other hand, line profile differences between some
components may be indicative of microlensing, which is expected to
magnify the --small-- continuum region and not the --larger-- broad
emission line region.

We first note that the line profiles in components A and B are
essentially identical up to the scaling factor. This suggests that
neither A nor B are strongly affected by microlensing.  The small
difference observed in the [\ion{O}{iii}] lines may be related to the
fact that the narrow line region could be partially resolved (Chantry
and Magain \cite{cha07}).  The wavelength dependence of the scaling
factor reveals higher dust extinction along the B line of sight than
along the A one, as discussed in detail in Sect.~\ref{sec:exti}. In
the following we consider the average spectrum of components A and B,
denoted AB, as the reference spectrum unaffected by microlensing
effects. As seen in the upper panel of Fig.~\ref{fig:specvis}, the
spectrum of \obj\ changes regularly with time, suggesting intrinsic
variations in the quasar outflow (this is further discussed in
Sect.~\ref{sec:vari}). These changes are observed in all components,
indicating that the time scale of the intrinsic line profile variation
is longer than the time delays between the four images. In fact, the
longest time delay is not larger than one month according to the
observations of Goicoecha and Shalyapin (\cite{goi10}) and in
agreement with models\footnote{We also estimate small time delays,
typically less than 10 days, considering a classical singular
isothermal ellipsoid (SIE)+shear model, a SIE+shear+galaxy model
proposed by MacLeod et al. (\cite{mac09}), as well as using pixellated
models with a symmetric mass distribution (Saha and Williams
\cite{sah04}).}  (Kayser et al. \cite{kay90}, Chae and Turnshek
\cite{cha99}).

The spectrum of component D is clearly different when compared to AB.
After scaling the continua, the emission appears less intense in D.
This behavior is observed at all epochs and in the different spectral
lines, superimposed onto the intrinsic time variations seen in all
components. This is a clear signature of a long-term microlensing
effect in D with amplification of the continuum with respect to the
emission lines. The timescale of the effect is in agreement with
previous estimates, i.e. on the order of 10 years (e.g. Hutsem\'ekers
\cite{hut93}).  As first pointed out by Angonin et al. (\cite{ang90}),
a difference is also observed in the absorption profiles. This
difference is especially strong in the 1989 and 1993 spectra.  This is
a priori not expected since the region at the origin of the observed
absorption lines has the same spatial extent as the continuum
source. Differential microlensing of absorbing clouds smaller in
projection than the continuum source has been proposed (Angonin et al.
\cite{ang90}). However the timescale of such an event is expected to
be smaller than 1 year (Hutsem\'ekers \cite{hut93}), ruling out this
interpretation. Instead, we interpret the difference in the absorption
profiles as due to the superposition of an emission line onto a nearly
black absorption profile (Sect.~\ref{sec:micbal}).

Although not as strong as in component D, spectral differences are
also observed when comparing C to AB. After scaling the continua, the
emission lines appear slightly stronger in C at all epochs, with a
possible differential effect between the broad H$\beta$ and the
narrower [\ion{O}{iii}] lines.  This suggests that microlensing also
affects component C, de-amplifying the continuum with respect to the
emission lines.

The scaling factor used to fit the continuum of D to that one of AB
clearly depends on wavelength.  This can be explained either by
chromatic microlensing (the source of UV continuum is less extended
than the source of visible continuum and then more magnified),
differential extinction (the extinction along the AB line of sight is
higher than along the line of sight to D), wavelength-dependent
dilution of the quasar continuum by the host galaxy light, or a
combination thereof.  Such a strong wavelength dependence of the
scaling factor is not observed when comparing C to AB.  The origin of
this effect is further discussed in Sect.~\ref{sec:mlconti}.

\subsection{The extinction curve from A/B}
\label{sec:exti}

\begin{figure}[t]
\resizebox{\hsize}{!}{\includegraphics*{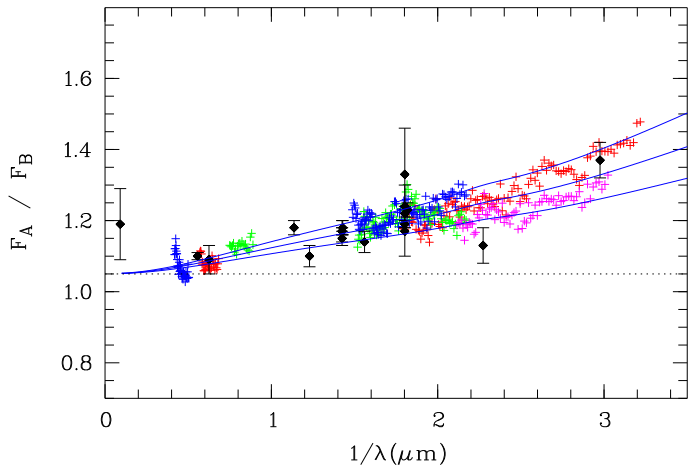}}
\caption{The flux ratio $F_{\rm A} / F_{\rm B}$ from the UV-visible
spectra of 1989, 1993, 1994, 2000, and from the near-infrared spectra
of 2005 (crosses of different colors). Photometric data point obtained
at different epochs are superimposed (black diamonds with error
bars). The continuous lines represent SMC-like extinction curves
redshifted to $z_l = 1.0$.}
\label{fig:ratio}
\end{figure}

Since images A and B are not significantly affected by microlensing,
the wavelength dependence of their flux ratio can be interpreted in
terms of differential extinction in the lensing
galaxy. Fig.~\ref{fig:ratio} illustrates the observed flux ratio
$F_{\rm A} / F_{\rm B}$ using all available spectroscopic data
(slightly filtered and smoothed). Photometric data are
superimposed. They were collected from Angonin et al. (\cite{ang90}),
{\O}stensen et al. (\cite{ost97}; the ratios $F_{\rm A} / F_{\rm B}$
are averaged per filter), Turnshek et al. (\cite{tur97}), Chae and
Turnshek (\cite{cha01}), Kneib et al. (\cite{kne98}), Chantry and
Magain (\cite{cha07}), MacLeod et al. (\cite{mac09}).

There is a lot of dispersion in the measured flux ratios which arises
not only from inacurracies in the data but also from a possible
combined effect of intrinsic photometric variations and time delay
(cf.  {\O}stensen et al. \cite{ost97}).  A clear trend is nevertheless
observed, indicating higher extinction along the line of sight to
image B, as suggested by Turnshek et al. (\cite{tur97}). The flux
ratio is reasonably well modeled using \begin{eqnarray} \frac{F_{\rm
A}}{F_{\rm B}} = \frac{F_{\rm A0}}{F_{\rm B0}} \,\, 10^{-0.4 \, \xi (
\lambda ) \, \Delta A_B } \end{eqnarray} where $\Delta A_B = A_B ({\rm
A}) - A_B ({\rm B})$ is the difference of extinction between the A and
B line of sights measured in the $B$ filter and $\xi(\lambda)$ the
extinction curve tabulated in Pei (\cite{pei92}), redshifted to the
lens redshift $z_l \simeq 1.0$ (Kneib et al. \cite{kne98}). Since no
obvious 2200\AA\ feature typical of the Milky Way extinction is
observed at 2.3 $\mu$m$^{-1}$ (Fig.~\ref{fig:ratio}), we adopt a
SMC-like extinction curve. A reasonably good fit is obtained with
$F_{\rm A0} / F_{\rm B0}$ = 1.05$\pm$0.02 and $\Delta A_B$ =
$-$0.09$\pm$0.02 (Fig.~\ref{fig:ratio}).  With $z_l \simeq 1.88$
(Goicoecha and Shalyapin \cite{goi10}) a similar fit is obtained with
$\Delta A_B$ = $-$0.055$\pm$0.015.  For $z_l \simeq$ 1.0
(1.88), the lines of sight to images A and B sample regions of the
lens galaxy located at respectively 4.9 (5.2) kpc and 6.0 (6.4) kpc
from the galaxy center, adopting the astrometry of Chantry and Magain
(\cite{cha07}) and assuming a flat cosmology with $\Omega_m$ = 0.27
and H$_{\rm 0}$ = 70 km s$^{-1}$ Mpc$^{-1}$.  The flux ratio $F_{\rm
A0} / F_{\rm B0}$ = 1.19$\pm$0.10 measured by MacLeod et
al. (\cite{mac09}) at 11 $\mu$m in the mid-infrared, i.e. at
wavelengths were both extinction and microlensing are expected to be
negligible, is compatible with the extinction corrected flux ratio
$F_{\rm A0} / F_{\rm B0}$ we derive, although marginally higher.

\subsection{Intrinsic line profile variations}
\label{sec:vari}

Time variations in the absorption line profiles of BAL quasars are not
uncommon (Barlow et al. \cite{bar89,bar92}, Gibson et
al. \cite{gib08}).  In \obj , Turnshek et al. (\cite{tur88}) reported
a deepening of the \ion{Si}{iv} BAL between 1981 and 1985.  On the
contrary, between 1989 and 2005 (Fig.~\ref{fig:specvis}), variations
are observed as a gradual decrease of the depth of the BAL
high-velocity part, the deepest component of the profile being
essentially unaffected.  Variations appear more complex in
\ion{Si}{iv} than in \ion{C}{iv}, affecting a larger part of the
absorption profile (see also Fig.~\ref{fig:sepavis}).

The strongest change occurs between 1989 and 1993 and corresponds to
an increase of the luminosity (Remy et al. \cite{rem96}, {\O}stensen
et al. \cite{ost97}). Moreover, stronger absorption is accompanied by
stronger emission, which is an indication that resonance line
scattering can play an important role in the emission line formation.
In the \ion{C}{iv} line of the AB spectrum, the high-velocity
absorption appears $\sim$15\% larger in 1989 than in 2005 while the
emission is $\sim$25\% more intense.  In the framework of resonance
scattering where each absorbed photon is re-emitted, this may suggest
that the high-velocity outflow has more scattering material
perpendicular to the line of sight than absorbing material along the
line of sight. This also requires a large covering factor.

\section{Decomposition of the line profiles}
\label{sec:decomp}

\subsection{The method}
\label{sec:method}

We follow a method similar to that used in Sluse
et~al.~(\cite{slu07}).  Assuming that the observed spectra $F_i$ of
the different images are made of a superposition of a spectrum $F_M$
which is only macrolensed and of a spectrum $F_{M\mu}$ which is both
macro- and microlensed, it is possible to extract the components $F_M$
and $F_{M\mu}$ by using pairs of observed spectra. Indeed, considering
a line profile, we can write
\begin{eqnarray} F_1 & = & M F_M + M \mu F_{M\mu}\\ F_2 & = & F_M +
F_{M\mu} \; .
\end{eqnarray} 
where $M=M_1/M_2$ is the macro-amplification ratio between images~1
and~2 and $\mu$ the micro-amplification factor of image~1. We assume
image 2 not affected by microlensing.  To extract $F_M$ and $F_{M\mu}$
when $M$ is not known a priori, these equations can be conveniently
rewritten
\begin{eqnarray} 
F_M \ & = & \frac{-A \;}{A - M} \; \; \left( \frac{F_1}{A} - F_2 \right) 
\\ 
F_{M\mu}
& = & \frac{M}{A - M } \; \; \left( \frac{F_1}{M} - F_2 \right) 
\end{eqnarray} 
where $A = M\mu$.  Considering the line profiles as the sum of a
continuum (absorbed or not in the blue) and an emission profile, we
write $F_2 = F_c + F_e \,$, $F_1 = F'_{c} + F'_{e} \,$,  and then
\begin{eqnarray} 
F_M \ & = & \frac{-A \;}{A - M} \; \; \left( \frac{F'_{c}}{A} - F_c + \frac{F'_{e}}{A} - F_e \right) 
\\ 
F_{M\mu}
& = & F_c + \frac{M}{A - M } \; \; \left( \frac{F'_{e}}{M} - F_e \right) \; .
\end{eqnarray} 
Up to a scaling factor,  $F_M$ only depends on $A$, while
$F_{M\mu}$ only depends on $M$ (Eqs.~4 and 5). $A$ is the scaling
factor between the $F_1$ and $F_2$ continua: $A = F'_{c} / F_c$. It
can be accurately determined as the value for which $F_M(A) = 0$ in
the continuum adjacent to the line profile, where $F_e = F'_{e} =
0$~(Eq.~6).  Assuming $A$ constant across the line profile, $F_M$ then
only contains a contribution from the emission profile.  Under the
assumption that at least a portion of the observed emission profile is
only macro-amplified (i.e. not micro-amplified), the
macro-amplification factor can be estimated as the value of $M$ such
that $F_{M\mu}(M) = F_c$ over that part of the line profile (Eq.~7,
see also Appendix A).  The micro-amplification factor of the continuum
is then estimated at the wavelength of the line profile: $\mu =
A/M$. On the other hand, if the emission line is amplified exactly
like the continuum, microlensing cannot be distinguished from
macrolensing.

Eqs.~6 and 7 show that if the emission profile is only macro-amplified
(i.e. if $F'_{e} = M \, F_{e}$), $F_{M\mu}$ only contains the
underlying continuum $F_c$, while $F_M$ contains the full emission
profile $F_e$.  If the emission profile is micro-amplified to some
extent, parts of it will appear in both $F_{M\mu}$ and $F_M$ (see
Appendix A).  More specifically, $(F_{M\mu}-F_c)$ can be seen as the
part of the emission profile which is not only macro-amplified, and
$F_M$ as the part of the emission profile which is not amplified like
the continuum.

The micro-amplification factor $\mu$ and the macro-amplification
factor $M$ possess some specific chromatic behaviors.  While $\mu$ is
assumed constant over the small wavelength range spanned by a line
profile, it can be different at the wavelengths corresponding to
different line profiles. Indeed, the micro-amplification of the
continuum is related to the effective size of the continuum source
which can be wavelength-dependent.  The macro-amplification factor $M$
may also contain a wavelength-dependent contribution due to
differential extinction in the lensing galaxy, since extinction, like
macrolensing, acts on the line profile as a whole.  Finally,
concerning the time dependence properties, $M$ is expected to remain
identical at different epochs of observation while $\mu$ can be
time-dependent.

\subsection{The results}
\label{sec:results}

\subsubsection{Line profile decomposition from the (D,AB) pair}

\begin{table}[t]
\caption{Amplification factors determined from the (D,AB) pair}
\label{tab:ampli}
\begin{tabular}{lcccc}\hline\hline \\[-0.10in]
   Lines &  Date &  $A$ & $M$ & $\mu$\\ 
\hline \\[-0.10in]
 \ion{Si}{iv}-\ion{C}{iv} & 1989 & 0.645  $\pm$0.015 & 0.390 $\pm$0.020 & 1.65  $\pm$0.12 \\
 \ion{Si}{iv}-\ion{C}{iv} & 1993 & 0.915  $\pm$0.015 & 0.460 $\pm$0.040 & 1.99  $\pm$0.21 \\
 \ion{Si}{iv}-\ion{C}{iv} & 2000 & 0.775  $\pm$0.010 & 0.400 $\pm$0.020 & 1.94  $\pm$0.12 \\
 \ion{Si}{iv}-\ion{C}{iv} & 2005 & 0.775  $\pm$0.010 & 0.400 $\pm$0.040 & 1.94  $\pm$0.22 \\
 Ly$\alpha$-\ion{N}{v}    & 1994 & 1.050  $\pm$0.025 & 0.460 $\pm$0.030 & 2.28  $\pm$0.20 \\
 Ly$\alpha$-\ion{N}{v}    & 2000 & 0.805  $\pm$0.015 & 0.430 $\pm$0.030 & 1.87  $\pm$0.17 \\
 H$\beta$-[\ion{O}{iii}]  & 2005 & 0.630  $\pm$0.010 & 0.425 $\pm$0.015 & 1.48  $\pm$0.08 \\
 H$\alpha$                & 2005 & 0.540  $\pm$0.015 & 0.395 $\pm$0.015 & 1.37  $\pm$0.09 \\
 \hline\\[-0.2cm]
\end{tabular}
\end{table}

\begin{figure}[t]
\resizebox{\hsize}{!}{\includegraphics*{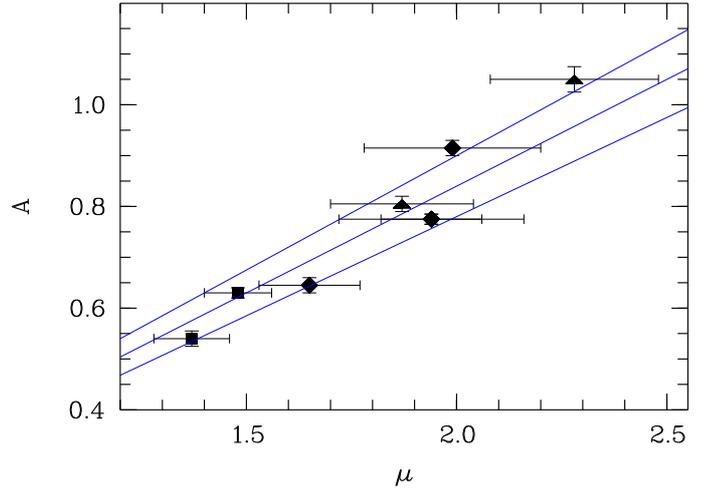}}
\caption{The relation between $A$ and $\mu$ measured from the spectral
analysis of the (D,AB) pair (Table~\ref{tab:ampli}). Diamonds
represent data from \ion{Si}{iv}--\ion{C}{iv} lines, triangles data
from Ly$\alpha$+\ion{N}{v} lines, and squares data from the Balmer
lines. The continuous lines represent $A = M \mu$ where $M$ = 0.39,
0.42, and 0.45.  }
\label{fig:ampli}
\end{figure}

We extract $F_{M}$ and $F_{M\mu}$ from the spectra of \obj\ using $F_1
= F_{\rm D}$ and $F_2 = F_{\rm AB}$ in Eqs. 4 and 5.  We have $M < 1$
and $\mu > 1$ since D is fainter than AB and its continuum amplified
(Sect. \ref{sec:microe}).  $A$ is computed as the value for which
$F_M(A) = 0$ in the continuum windows adjacent to the line profiles
(cf. Sect.~\ref{sec:microe}). Within the uncertainties, $A$ is the
inverse of the scaling factor determined in Figs.~\ref{fig:specvis}
and~\ref{fig:specuvir}.  $M$ is computed as the largest value for
which $F_{M\mu}(M) \geq F_c$ over the whole line profile.  For the
resonance line profiles, the continuum may be completely absorbed at
some wavelengths such that $F_{M\mu}(M)$ must also be larger than
zero.  Measurements of $A$ and $M$ are given in Table~\ref{tab:ampli}
with $\mu = A / M$.  Taking into account the noise and possible
contaminating features in the observed spectra, a range of acceptable
values is obtained which provides a rough estimate of the
uncertainties. These are typically 2--3\% for $A$ and 4--10\% for $M$.
As expected, the macro-amplification factor $M$ is independent of the
epoch of observation, the dispersion of the values being in agreement
with the uncertainties.  The temporal variation of $A$ then
essentially comes from the variation of the micro-amplification of the
continuum. This is shown in Fig.~\ref{fig:ampli}. This figure also
suggests that, within the uncertainties, there is no significant
wavelength dependence of $M$ due to a differential reddening between
AB and D (see also Sect.~\ref{sec:macrol}).

\begin{figure*}[!]
\resizebox{17cm}{!}{\includegraphics*{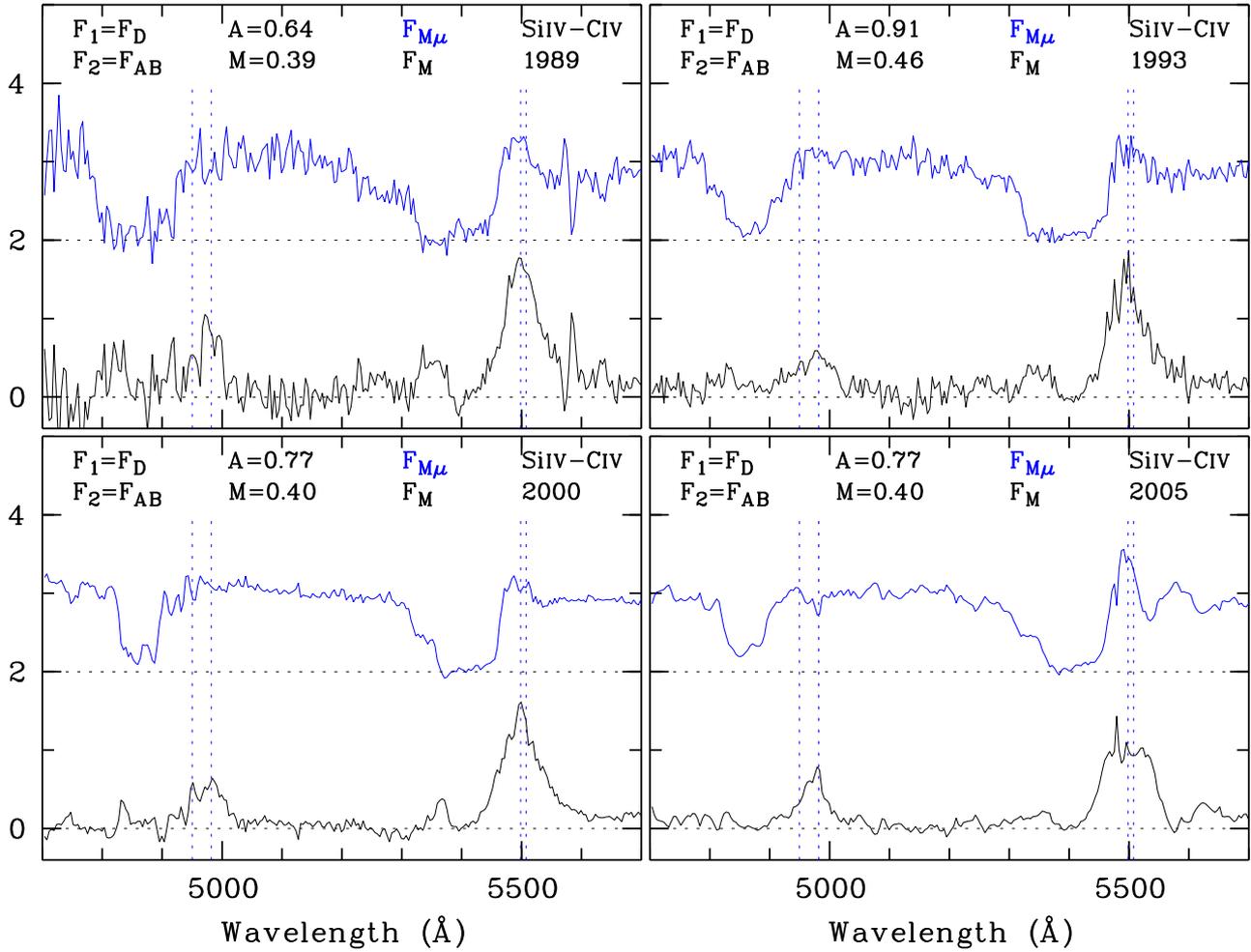}}
\caption{The microlensed $F_{M\mu}$ and macrolensed-only $F_{M}$
spectra of \obj\ extracted from the comparison of the D and AB
spectra, at different epochs.The \ion{Si}{iv} and \ion{C}{iv} line
profiles are illustrated.  The amplification factors $A$ and $M$ are
given for each epoch.  Ordinates are relative fluxes, the $F_{M\mu}$
spectrum being shifted upwards by 2 units.  Vertical dotted lines
indicate the positions of the spectral lines at $z = 2.553$. }
\label{fig:sepavis}
\end{figure*}

\begin{figure*}[!]
\resizebox{17cm}{!}{\includegraphics*{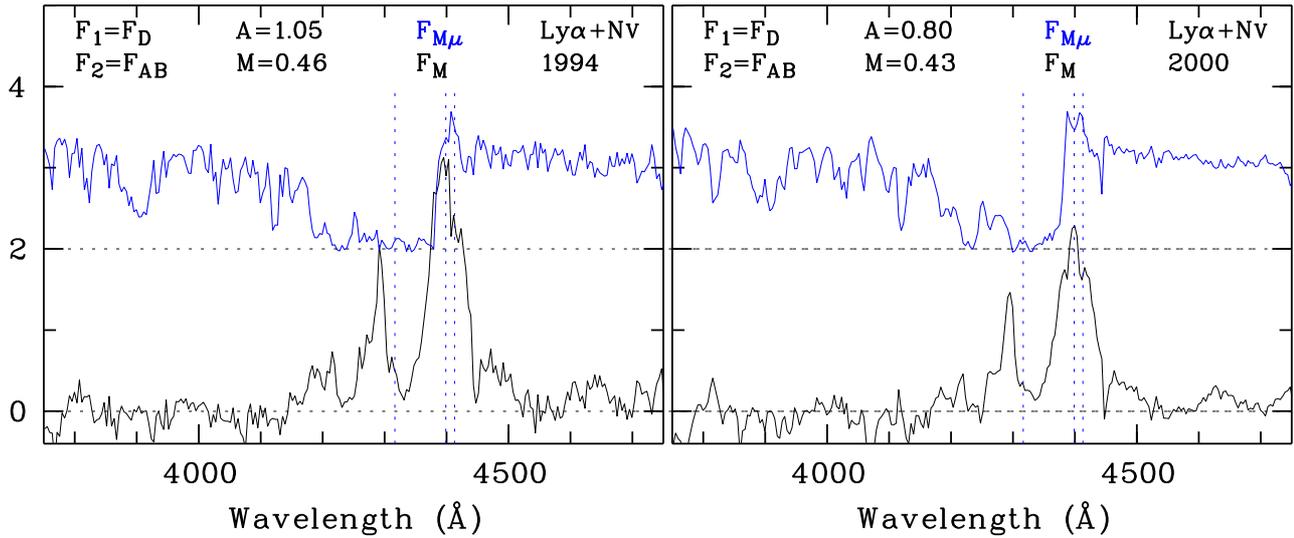}}
\caption{Same as Fig.~\ref{fig:sepavis}, but for the Ly$\alpha$ +
\ion{N}{v} line profile.}
\label{fig:sepauv}
\end{figure*}

\begin{figure*}[!]
\resizebox{17cm}{!}{\includegraphics*{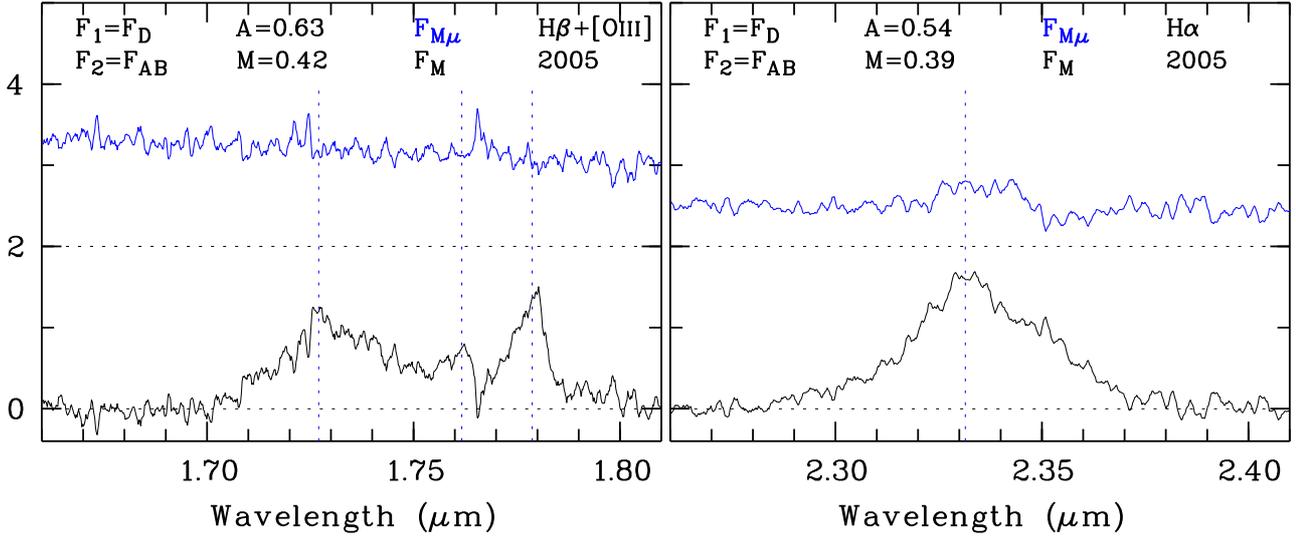}}
\caption{Same as Fig.~\ref{fig:sepavis}, but for the H$\beta$ +
[\ion{O}{iii}] (left)
and H$\alpha$ (right) line profiles.}
\label{fig:sepair}
\end{figure*}

With the measured $A$ and $M$, we then compute $F_{M}$ and $F_{M\mu}$
from Eqs.~4 and~5.  The results of the line profile decompositions are
given in Figs.~\ref{fig:sepavis} to~\ref{fig:sepair}. Over large parts
of the emission profile (mainly in the wings), $F_{M\mu} = F_c$. This
indicates that at least a part of the observed emission line profile
is unaffected by microlensing and that $M$ actually represents the
macro-amplification factor. We emphasize that the
derived $F_{M}$ and $F_{M\mu}$ profiles are robust against small
changes of $A$ or $M$.

Fig.~\ref{fig:sepavis} illustrates the spectral decomposition for the
\ion{Si}{iv} -- \ion{C}{iv} region.  Although different values of $A$
(or $\mu$) are found at different epochs, the extracted spectra are
remarkably consistent. The microlensed part of the spectrum,
$F_{M\mu}$, contains the continuum with the full absorption profiles
as well as a small contribution from the core (not the wings) of the
emission profiles. The bulk of the emission lines appears in the
macrolensed-only part of the spectrum $F_{M}$, clearly showing a
two-peak structure in \ion{C}{iv} (at 5350~\AA\ and 5500~\AA ).
Recall that $F_M$ shows the flux emitted from a large region of the
quasar (much larger than the Einstein radius of the microlens) whereas
$F_{M\mu}$ shows the flux emitted from a smaller region (comparable to
and smaller than the Einstein radius).  Interestingly enough, the
small part of the emission profile observed in $F_{M\mu}$ is the core,
i.e. the low-velocity part.  The temporal variations of the absorption
are particularly well seen in the $F_{M\mu}$ spectrum of \ion{Si}{iv}.

\begin{figure*}[!]
\resizebox{17cm}{!}{\includegraphics*{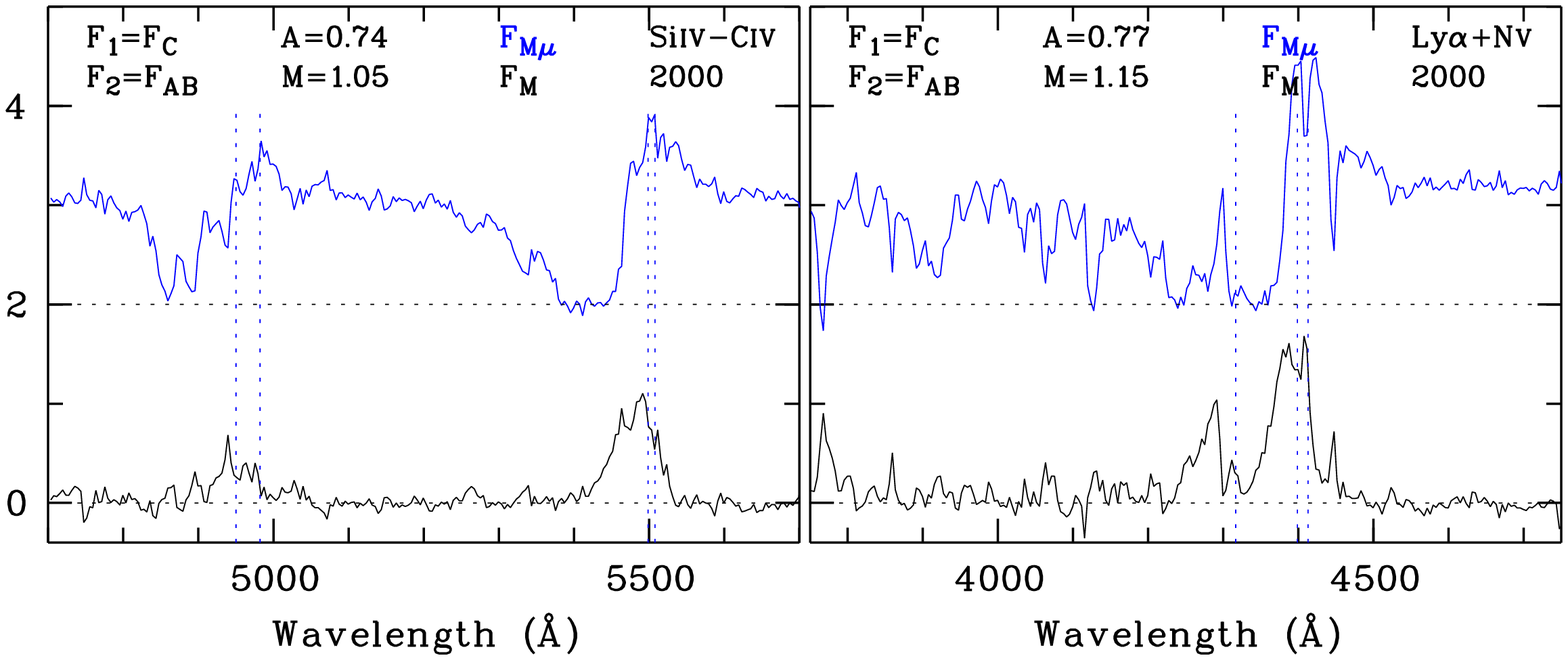}}
\caption{Same as Fig.~\ref{fig:sepavis} and ~\ref{fig:sepauv} but the
microlensed $F_{M\mu}$ and macrolensed-only $F_{M}$ spectra are
extracted from the comparison of the C and AB spectra obtained in
2000.  The \ion{Si}{iv} and \ion{C}{iv} (left) and Ly$\alpha$ +
\ion{N}{v} (right) line profiles are illustrated.}
\label{fig:sepac}
\end{figure*}

The rest-frame UV spectra, and more particularly the Ly$\alpha$ +
\ion{N}{v} region, are similarly analyzed (Fig.~\ref{fig:sepauv}).
Although the decomposition is less accurate due to structures in the
continuum blueward of Ly$\alpha$ (possibly due to narrow absorption
features and inaccuracies in the wavelength calibration), the
extracted spectra $F_{M\mu}$ and $F_{M}$ show the same qualitative
behavior as observed in the \ion{C}{iv} and \ion{Si}{iv} line
profiles.

Fig.~\ref{fig:sepair} shows the decomposition for the H$\beta$ +
[\ion{O}{iii}] and H$\alpha$ line profiles.  The micro-amplified
spectrum is clearly a flat continuum in the H$\beta$ + [\ion{O}{iii}]
spectral region while there is some evidence that the core (and not
the wings) of the H$\alpha$ emission line is micro-amplified.

\subsubsection{Line profile decomposition from the (C,AB) pair}

The same kind of analysis can be done using the pair (C,AB). However
the line profile differences are more subtle
(cf. Figs.~\ref{fig:specvis} and~\ref{fig:specuvir}) so that the value
of $M$ is closer to the value of $A$ (i.e.  $\mu$ closer to~1) in
Eqs.~4 and~5, making the extracted spectra noisier.  Only the spectra
obtained in 2000, which show the most conspicuous profile differences,
are considered here.  A de-magnification of the continuum explains the
observations (Sect.~\ref{sec:microe}) and the resulting $F_{M}$ and
$F_{M\mu}$ are illustrated in Fig.~\ref{fig:sepac}.  They are roughly
similar to those derived from the pair (D,AB) but the part of the
emission profile which is micro-amplified is different. While only a
small part of the core of the emission is seen in the $F_{M\mu}$
profile derived from the (D,AB) pair, the red wing of the emission
profile is also observed in $F_{M\mu}$ computed from (C,AB).  This red
wing and the blue emission peak are not seen in $F_{M}$, suggesting
that the high-velocity component of the \ion{C}{iv} resonance line is
micro-deamplified like the continuum.  This is not unexpected since,
for a given Einstein radius, demagnification regions with relatively
smooth $\mu$ variations can extend on larger scales than amplification
regions (e.g. Lewis and Ibata \cite{lew04}).  The emission line core,
which appears in both $F_{M\mu}$ and $F_{M}$, should originate, at
least in part, from a region more extended than the high-velocity
component.

From the 2005 near-infrared spectra (the decomposition of which is not
shown), we measure $A$ = 0.74 for both the H$\beta$+[\ion{O}{iii}] and
H$\alpha$ regions. The condition $F_{M\mu}(M) \geq F_c$ is verified at
the wavelengths of H$\beta$ or H$\alpha$ with $M \simeq$ 0.87, while
$F_{M\mu}(M) \geq F_c $ is obtained at the wavelength of the
[\ion{O}{iii}] lines with $M \simeq$ 1.03. The latter value is
comparable to the value of $M$ derived from the UV-visible resonance
lines.  Although the narrow line region may be partially resolved
(Chantry and Magain \cite{cha07}), the [\ion{O}{iii}] emission lines
are expected to originate from a larger region and then less affected
by microlensing, making $M \simeq$ 1.03 a more plausible estimate.  In
this case, with $\mu = A/M \simeq$ 0.72 for the micro-amplification
factor of the continuum, the Balmer emission lines do appear in both
$F_{M\mu}$ and $F_{M}$. This means that they are also
micro-deamplified although not as much as the continuum.

\section{Lensing in \obj\ }
\label{sec:lensing}

\subsection{The macro-amplification factors}
\label{sec:macrol}

In principle, the variation of $M$ with the wavelength can be
attributed to differential extinction.  Unfortunately, for the (D,AB)
pair, the wavelength dependence is not clear enough to extract an
extinction curve, given the uncertainties on the determination of $M$
(Fig.~\ref{fig:ampli} and Table~\ref{tab:ampli}).  The results
nevertheless suggest that the differential extinction between AB and D
is lower than between A and B for which the extinction at Ly$\alpha$
is $\sim$ 1.2 times the extinction at H$\alpha$
(Fig.~\ref{fig:ratio}). As a consequence, the value $M$(D,AB) =
0.395$\pm$0.015 determined at the wavelength of H$\alpha$, i.e. in the
reddest part of our spectra, should not differ from the true
macro-amplification factor by more than~2\%.

For the (C,AB) pair, we conservatively adopt $M$(C,AB) = 0.95 $\pm$
0.08 at the wavelength of H$\alpha$.  Comparing with $M$(C,AB) = 1.15
$\pm$ 0.05 and $M$(C,AB) = 1.05 $\pm$ 0.05 at the wavelengths of
Ly$\alpha$ and \ion{C}{iv}, respectively (Fig.~\ref{fig:sepac}),
$M$(C,AB) might be slightly wavelength dependent, providing marginal
evidence that extinction is lower for C than for AB. Since the
differential extinction remains low, we also assume that it does not
affect the macro-amplification factor determined at the wavelength of
H$\alpha$ by more than~2\%.

The flux ratios with respect to component A are then $F_{\rm
B0}$/$F_{\rm A0}$ = 0.95$\pm$0.02 (Fig.~\ref{fig:ratio}), $F_{\rm
C0}$/$F_{\rm A0}$ = 0.93$\pm$0.10 and $F_{\rm D0}$/$F_{\rm A0}$ =
0.39$\pm$0.04. The fact that $F_{\rm C0}$/$F_{\rm A0}$ and $F_{\rm
D0}$/$F_{\rm A0}$ are different from the values estimated from
photometry is due to a significant de-amplification of the C continuum
and to a significant amplification of the D continuum, as derived from
the analysis of the spectral lines. This emphasizes the need to
properly correct for microlensing before interpreting the flux ratios.

MacLeod et al.~(\cite{mac09}) determined $F_{\rm B0} / F_{\rm A0}$ =
0.84$\pm$0.07, $F_{\rm C0} / F_{\rm A0}$ = 0.72$\pm$0.07 and $F_{\rm
D0} / F_{\rm A0}$ = 0.40$\pm$0.06 at 11 $\mu$m in the mid-infrared,
i.e., where microlensing and extinction are thought to be negligible.
Although only marginally different, the flux ratios of the B and C
components relative to A seem slightly smaller than ours.  If real,
the origin of such a discrepancy is not clear but could be related to
the intense starburst activity detected in the host galaxy of \obj\
(Lutz et al. \cite{lut07}, Bradford et al. \cite{bra09}), which
possibly contaminates with PAH emission the 11.2 $\mu$m (3.2 $\mu$m
rest-frame) flux measurements.

\subsection{Microlensing of the continuum source}
\label{sec:mlconti}

\begin{figure}[t]
\resizebox{\hsize}{!}{\includegraphics*{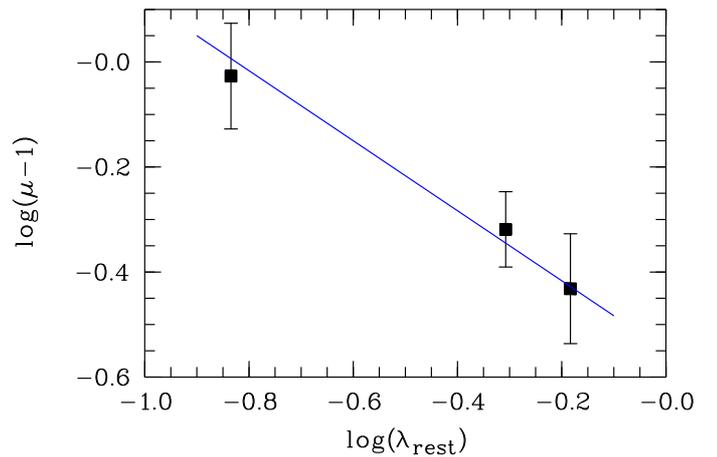}}
\caption{The micro-amplification factor $\mu$ measured in 2005 (Table
\ref{tab:ampli}) against the wavelength of observation expressed in
the quasar rest-frame (in $\mu$m).  The straight line represents the
model prediction (see text).}
\label{fig:chroml}
\end{figure}

The micro-amplification factor $\mu$ determined in component D depends
on both the date and the wavelength (Table~\ref{tab:ampli}).

Considering the \ion{Si}{iv}-\ion{C}{iv} spectral region, the
strongest variation of $\mu$ occurs between 1989 and 1993 (see
also Figs. \ref{fig:specvis} and \ref{fig:sepavis}).  It roughly
corresponds to a relative photometric variation between A and D which
can be observed in the V light curves of \obj\ presented by Remy et
al. (\cite{rem96}) and {\O}stensen et al. (\cite{ost97}), superimposed
onto the common intrinsic variation of the 4 components.  Between 1993
and 2000, the variation of $\mu$ is weaker, in agreement with the HST
photometry in the F555W filter reported by Turnshek et
al. (\cite{tur97}) and Chae et al.  (\cite{cha01}). 

At a given epoch, $\mu$ decreases with increasing wavelength,
suggesting chromatic magnification of the continuum source.  This is
best seen in the 2005 data (obtained within a 2 month interval) which
span the largest wavelength range. We emphasize that $\mu$, when
determined from the line profiles, is not contaminated by differential
extinction (Sect.~\ref{sec:method}).  In Fig.~\ref{fig:chroml}, we
plot the values of $\mu$ as a function of the wavelength of
observation in the quasar rest-frame.

The magnification $\mu$ of an extended source close to a caustic can
be written
\begin{eqnarray}
\mu = \mu_0 \, + \frac{g}{\sqrt{R_S/R_E}} \; \zeta(d)
\end{eqnarray} 
where $R_S$ is the source radius, $R_E$ the Einstein radius of the
microlens projected onto the source plane, $g$ is a constant on the
order of unity, $\zeta(d)$ a function which depends on the distance to
the caustic and $\mu_0$ a constant ``background'' magnification
(e.g. Schneider et al. \cite{sch92}, Witt et al. \cite{wit93}).  In
the framework a simple model where the continuum is emitted by a
Shakura-Sunyaev (\cite{sha73}) thin accretion disk thermally
radiating, $R_S(\lambda) \propto \lambda ^{4/3}$ (e.g. Poindexter et
al. \cite{poi08}) such that we may expect
\begin{eqnarray}
\log \, (\mu -\mu_0)\,  =  \, -\frac{2}{3} \log\, (\lambda) + C \, .
\end{eqnarray} 
As seen in Fig.~\ref{fig:chroml}, this model nicely reproduces the
data using $C$ = $-$0.55 and assuming for simplicity $\mu_0$ = 1
(i.e. no background (de-~)magnification), supporting the idea of
chromatic magnification of a continuum emitted by a Shakura-Sunyaev
accretion disk.

In principle, we could have used the Ly$\alpha$+\ion{N}{v} and the
\ion{C}{iii}] $\lambda$1909 emission lines, also present in the 2005
visible spectra, to measure $\mu$ at other wavelengths. Although
tentative estimates do agree with the observed trend, the quality of
the data is not sufficient to derive reliable values of $\mu$ at
these wavelengths, due to the insufficient spectral resolution in the
complex Ly$\alpha$+\ion{N}{v} region and to the fact that the
\ion{C}{iii}] line is truncated. Clearly, with better quality data, it
could be possible to separate $M$ and $\mu$ at other wavelengths using
additional line profiles and thus derive the temperature profile of the
accretion disk.

From the value of the constant $C$, we can derive a rough estimate of
the size of the continuum source : $R_S(\lambda) / R_E \lesssim \,
\lambda^{4/3} \, 10^{-2C}$. The Einstein radius is computed to be $R_E
\simeq$ 0.01 $\sqrt{M/M_{\odot}}$ pc using $z_{l}$ = 1.0, a flat
cosmology, $\Omega_m$ = 0.27 and H$_{\rm 0}$ = 70 km s$^{-1}$
Mpc$^{-1}$. $M$ is the mass of the microlens.  Then, $R_S(\lambda)
\lesssim$ 0.012 $\sqrt{M/M_{\odot}}$ pc at the rest-frame UV
wavelength $\lambda$ = 0.15 $\mu$m. This limits tightens to
$R_S(\lambda) \lesssim$ 0.007 $\sqrt{M/M_{\odot}}$ pc with $z_{l}$ =
1.88.  These values agree with the radii obtained for the lensed
quasars HE1104$-$1805 and Q2237$+$0305 on the basis of their
photometric variability due to microlensing (e.g. Poindexter et
al. \cite{poi08}, Eigenbrod et al. \cite{eig08}, Anguita et
al. \cite{ang08b}).

Dilution of the quasar continuum --microlensed-- by the host
galaxy light --not microlensed-- can affect the interpretation of
the wavelength dependence of $\mu$.  Denoting $\alpha = F_{c}({\rm
host})/F_{c}({\rm qso})$ the ratio of the host and quasar continua at
a given wavelength, we find that the micro-amplification factor of the
quasar continuum $\mu_{q}$ is related to the measured $\mu$ by
\begin{eqnarray} 
\mu_{q} = \mu \, (1+\alpha) - \alpha
\end{eqnarray} 
if the host galaxy is unresolved and macrolensed as the quasar
(i.e. if $F'_{c} = M \, \mu_q \, F_{c}({\rm qso}) + M F_{c}({\rm
host})$), or by
\begin{eqnarray} 
\mu_{q} = \mu \, (1+\alpha) - \alpha/M
\end{eqnarray} 
if the host galaxy is resolved and not macrolensed (i.e. if $F'_{c} = M
\, \mu_q \, F_{c}({\rm qso}) + F_{c}({\rm host})$ assuming the host
contained in the measurement aperture).  Dilution by the host galaxy
is expected to be stronger in the rest-frame visible-infrared than in
the ultraviolet.  In the rest-frame visible of a luminous
high-redshift object, $\alpha \lesssim 0.15$ (e.g. Schramm et
al. \cite{sch08}) so that $\mu_q$ does not differ from $\mu$ by more
than 5\% (even less if the host is partially resolved). This is
negligible and within the uncertainties.

\subsection{Microlensing of a scattering region ?}

Chae et al. (\cite{cha01}) have obtained the first polarization
mesurements of the four images of \obj\ using the HST. The F555W
filter was used.  They noted that, in March 1999, the polarization
degree of component D might be higher that the polarization degree of
the other components (Table~\ref{tab:pola}). From this result, they
suggested that microlensing also affects the scattering region.  The
measurements obtained in June 1999 possibly indicate an intrinsic
variation of the polarization observed in all components. Similar
variations have been reported by Goodrich and Miller (\cite{goo95}).

\begin{table}[t]
\caption{Polarimetry of the four images}
\label{tab:pola}
\begin{tabular}{lrrrrrrr}\hline\hline \\[-0.10in]
   &  A &  B  &  C & D & ABC\\ 
\hline \\[-0.10in]
$p$ & &  & & &  \\ \hline \\[-0.10in]
1999/03&1.6$\pm$0.5&2.3$\pm$0.5&1.8$\pm$0.5&2.9$\pm$0.5&1.8$\pm$0.3\\
 1999/06&  - & - & - &1.0$\pm$0.6&0.9$\pm$0.3\\
 2008/05& 1.4$\pm$0.1&2.4$\pm$0.1&1.2$\pm$0.1&2.0$\pm$0.1&1.65$\pm$0.04\\
\hline \\[-0.10in]
$\theta$ & & & & &  \\ \hline \\[-0.10in]
1999/03& 75$\pm$9&65$\pm$6&71$\pm$8& 102$\pm$5&70$\pm$4\\
1999/06&      - & - & - &103$\pm$18&87$\pm$11\\
2008/05& 72$\pm$2&79$\pm$1&69$\pm$3&96$\pm$2&75$\pm$1\\
\hline\\[-0.2cm]
\end{tabular}\\
\footnotesize{The polarization degree $p$ is given in percent and the
polarization position angle $\theta$ in degree, East of North. The
data obtained in 1999 are from Chae et al.~(\cite{cha01}).}
\end{table}

Taking advantage of an excellent seeing, we were able to measure the
polarization of the 4 components of \obj\ in the V filter, from the
ground. Our measurements are also reported in Table~\ref{tab:pola}.
Within the uncertainties, the polarization degree of components A, B
and C do agree with the March 1999 values of Chae et al.
(\cite{cha01}), while the polarization degree of component D does
not. Instead, we find that the difference between the polarization
degrees measured in A and C and those ones measured in B and D is
significant. The difference between A and B is especially intriguing
since we found no significant microlensing effect neither in A nor in
B (at least before 2005). Possible interpretations could involve the
polarization due to an extended scattering region resolved by the
macrolens (possibly in the host galaxy, see Borguet et
al. \cite{bor08}), or a differential polarization induced by aligned
dust grains in the lens galaxy.  More data are clearly needed to
correctly understand the meaning of these measurements.

\section{Consequences for the BAL formation}
\label{sec:bal}

\begin{figure*}[!]
\resizebox{17cm}{!}{\includegraphics*{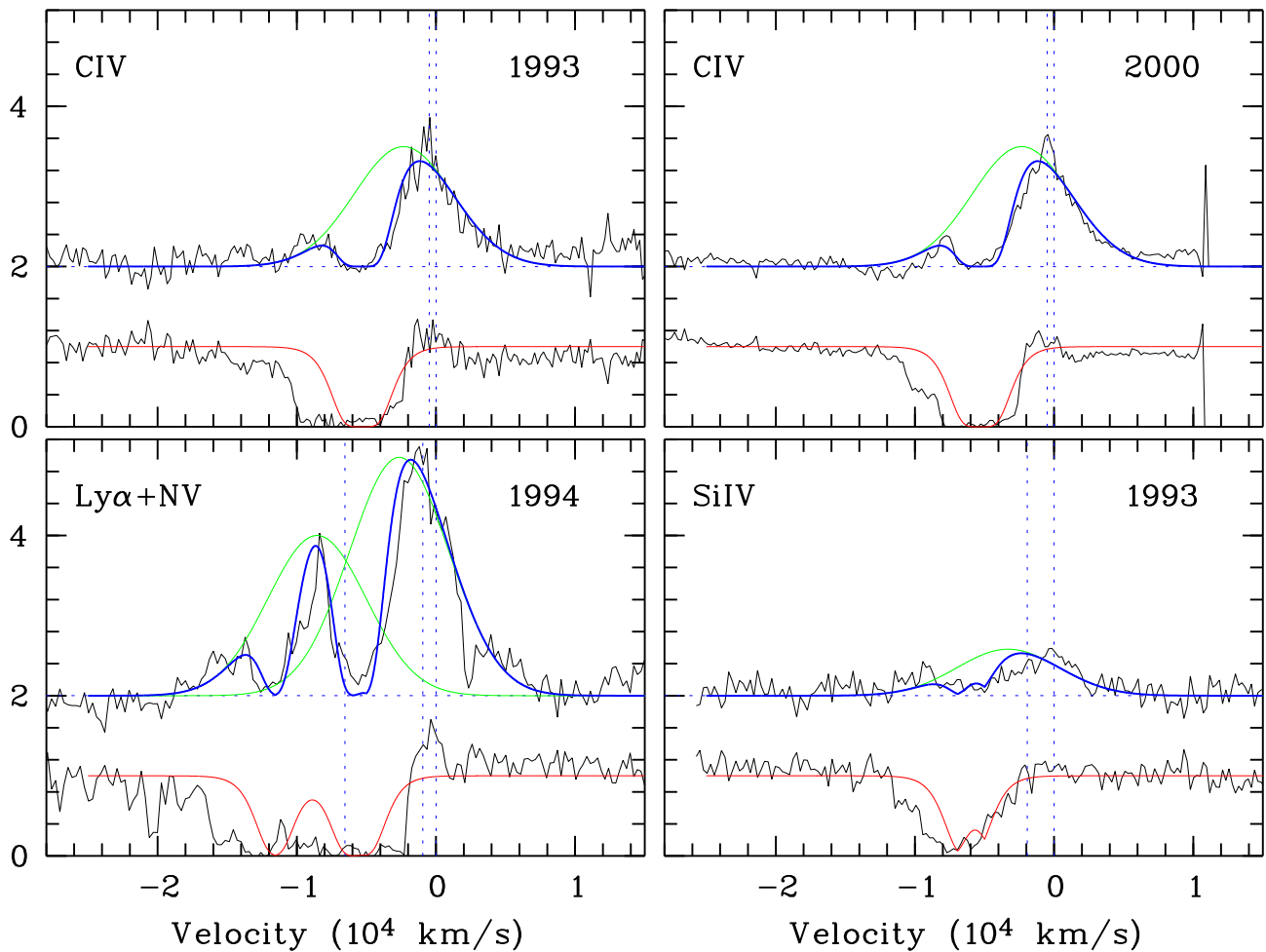}}
\caption{Selected $F_{M}$ (upper) and $F_{M\mu}$ (lower) spectra from
Figs.~\ref{fig:sepavis} and~\ref{fig:sepauv}, shown on a velocity
scale. The zero velocity corresponds to the red line of the doublets
redshifted using $z = 2.553$.  Ordinates are relative fluxes.  $F_{M}$
is shifted upwards by 2 units.  Vertical dotted lines indicate the
positions of the spectral lines. The profiles from the illustrative
model (Appendix B) are superimposed: the unabsorbed emission line
(green), the disk absorption profile (red) and the disk-absorbed
emission line (blue). Double-peaked emission profiles are
produced using a disk absorption profile which is contained in the
observed full (polar+equatorial) absorption line profile.}
\label{fig:vel}
\end{figure*}

In the previous sections we derived a consistent picture of
microlensing in \obj , showing that the continuum of component D (or
more precisely all the regions of the quasar located in a cylinder of
diameter $\sim 2 \, R_E$ oriented along the line of sight and
containing the continuum source) is magnified with respect to the more
extended regions at the origin of the emission lines. This allowed us
to disentangle the absorption part of the BAL profiles, essentially
$F_{M\mu}$, from the emission part, essentially $F_{M}$
(Figs.~\ref{fig:sepavis} to \ref{fig:sepair}).  The observed profiles
are equal to the sum of the $F_{M\mu}$ and $F_{M}$ spectra (Eq.~3).
The separation is robust against the uncertainties of the
amplification factors. It is however not perfect since emission which
originates from regions close to the continuum source in projection
appears in $F_{M\mu}$.  Selected spectra are illustrated in
Fig.~\ref{fig:vel} on a velocity scale.

The absorption profile of the \ion{C}{iv} BAL appears nearly black
extending from {\sl v} $\simeq$ $-$2000 km s$^{-1}$ to {\sl v}
$\simeq$ $-$10000 km s$^{-1}$. It is especially interesting to note
that the flow does not start at {\sl v}~=~0 in the rest-frame defined
by the [\ion{O}{iii}] emission lines. The part of the profile between
$-$8000 and $-$10000 km s$^{-1}$ is clearly variable between 1993 and
2000, showing a smaller depth in 2000. At a given epoch (1993--1994),
the absorption appears stronger in \ion{N}{v} and weaker in
\ion{Si}{iv}; this is best seen in the velocity range $-$2000 to
$-$4000 km s$^{-1}$ and indicates an ionization dependence of the
optical depth.  The extracted emission profiles show a double-peaked
structure which extends to the blue as far as the absorption profile
does. The blue peak at $-$8000 km s$^{-1}$ appears much fainter than
the red peak at $-$1000 km s$^{-1}$.  The full emission profile
(represented by the green line in Fig.~\ref{fig:vel}) is roughly
centered on the onset velocity of the flow ($-$2000 km s$^{-1}$), thus
blueshifted with respect to the [\ion{O}{iii}] rest-frame. In fact,
the full absorption + emission line profile appears in a rest-frame
blueshifted by $-$2000 km s$^{-1}$ with respect to the rest-frame
defined by the [\ion{O}{iii}] emission lines.  H$\alpha$, on the other
hand, appears redshifted (Fig.~\ref{fig:specuvir}). Although not
clearly understood, these line shifts are common in quasars
(e.g. Corbin \cite{cor90}, McIntosh et al.  \cite{mci99}) and
particularly strong in BAL QSOs (Richards et al. \cite{ric02}), in
agreement with our observations.  Interestingly enough, Nestor
et al. (\cite{nes08}) found a deficit of intrinsic \ion{C}{iv} Narrow
Absorption Line (NAL) systems at outflowing velocities lower than
2000 km s$^{-1}$, possibly due to overionization close to the
accretion disk.

The shape of the emission suggests that it is occulted by a strong
absorber, narrower in velocity than the full absorption profile, and
emitting little by itself.  Very similar absorption and emission
profiles are produced in the outflow model of Bjorkman et
al. (\cite{bjo94}) proposed for early-type stars.  We build on this
model to interpret our observations. A toy model, detailed in Appendix
B, is used for illustrative purposes (a full radiative transfer
modeling is beyond the scope of the present paper; it is
presented in Borguet and Hutsem\'ekers, \cite{bor10}, where details on
the flow geometry are also given).  We assume that the outflow in
\obj\ is constituted of two components: a quasi-spherically symmetric
``polar'' outflow, and a denser disk seen nearly edge-on.  The
equatorial disk expands slower than the polar wind and partly covers
it.  The polar outflow produces typical P Cygni line profiles
constituted of the superposition of a deep absorption extending from
$-$2000 to roughly $-$10000 km s$^{-1}$ and a symmetric emission due
to resonantly scattered photons (e.g. Lamers and Cassinelli
\cite{lam99}).  This emission (assumed gaussian shaped for simplicity)
is centered on {\sl v} $\simeq$ $-$2000 km s$^{-1}$ and extends from
$-$10000 to $+$6000 km s$^{-1}$. Both the remaining continuum and the
emission from the polar wind are absorbed in the equatorial disk.  A
double-peaked emission line is then produced (Fig.~\ref{fig:vel}),
little emission being expected from the edge-on disk.  As we can see
from Fig.~\ref{fig:vel}, this simple model is able to reproduce the
main characteristics of the intrinsic emission line profiles extracted
from the microlensing analysis. Since the disk is expected to also
absorb the continuum, its absorption profile must be contained within
the total polar+equatorial absorption profile, as illustrated.
Variability of the polar outflow optical depth will generate
variations at the high velocity end of the absorption accompanied by a
change in the resonantly scattered emission, as observed
(Sect.~\ref{sec:vari}; see also Bjorkman et al., \cite{bjo94}, for
simulations). Note that two-component winds have good theoretical
grounds (e.g. Murray et al. \cite{mur95}, Proga and Kallman
\cite{pro04}) and are supported by many observations interpreted with
either disks or polar flows (cf. Sect.~\ref{sec:intro}).

Spectropolarimetric observations of \obj\ (e.g. Goodrich and Miller
\cite{goo95}, Lamy and Hutsem\'ekers \cite{lam04}) provide additional
evidence favoring this kind of scenario.  First, the polarization
angle rotates within the absorption line profiles, suggesting the
existence of at least two sources and/or mechanisms of polarization.
The polar outflow and the disk, expected to produce perpendicular
polarizations, can play this role, especially in the case of BAL QSOs
with P Cygni-type profiles (Goodrich \cite{goo97}, Hutsem\'ekers et
al. \cite{hut98}, Lamy and Hutsem\'ekers \cite{lam04}). Furthermore,
the absorption in the polarized spectrum is clearly narrower than the
absorption in the direct spectrum (this is best observed in Fig.~3 of
Goodrich and Miller, \cite{goo95}), supporting the existence of a
slowly expanding equatorial disk which absorbs the polar-scattered
flux.

\subsection{Microlensing in the BAL}
\label{sec:micbal}

The difference observed in the BAL profiles of images AB and D (best
seen in \ion{C}{iv}, Fig.~\ref{fig:specvis}) can also be interpreted
in the framework of this outflow model.  In 1989 and 1993, the
intrinsic absorption at 5330~\AA\ ($-$9000 km s$^{-1}$) due to the
high-velocity ``polar'' outflow is nearly black
(Fig.~\ref{fig:sepavis} and~\ref{fig:vel}). In classical P~Cygni line
profile formation (e.g. Lamers and Cassinelli \cite{lam99},
Hutsem\'ekers and Surdej \cite{hut90}), this absorption is partially
filled in with emission resonantly scattered at the same velocity (the
blueward peak of the intrinsic emission, not absorbed by the slower
disk, and seen in Figs.~\ref{fig:sepavis} and~\ref{fig:vel}). Since
the emission line and the absorbed continuum react differently to the
magnification by the microlens, a spectral difference is observed in
the high-velocity part of the BALs seen in AB and D. Later, in 2000
and 2005, the high-velocity part of the BAL profile is less optically
thick: the absorption is not as deep as in the nineties and the blue
wing of the resonantly scattered emission which fills in the
absorption is accordingly weaker (the blue emission peak appears
narrower or less intense in 2000 and 2005, Figs.~\ref{fig:sepavis}
and~\ref{fig:vel}). As a consequence, the microlens-induced spectral
difference observed in the high-velocity part of the BALs appears
smaller at these epochs.

\section{Conclusions}
\label{sec:conclu}

Using 16 years of spectroscopic observations of the 4 components of
the gravitationally lensed BAL quasar \obj , we derived the following
results.

\noindent $-$ The strength of the BAL profiles gradually decreases
with time in all components. This intrinsic variation is accompanied
by a decrease of the intensity of the emission.

\noindent $-$ The spectral differences observed in component D can be
attributed to a long-term microlensing effect, in agreement with previous
studies. This effect consistently magnifies the continuum source of
image D, leaving the broad emission line region essentially
unaffected. We also find that the continuum of component C is most
likely de-magnified, while components A and B are not affected by
microlensing. Differential extinction is found between A and B.

\noindent $-$ Using a simple decomposition method to separate the part
of the line profiles affected by microlensing from the part unaffected
by this effect, we were able to disentangle the intrinsic absorption
(affected) from the emission line profile (unaffected). Consistent
results are obtained for the different epochs of observation.

\noindent $-$ Considering the macro- and micro-amplification factors
estimated with this method, we obtain a coherent view of lensing in
\obj .  In particular, we show that microlensing of the D continuum
source has a chromatic dependence which is compatible with a continuum
emitted by a standard Shakura-Sunyaev accretion disk.

\noindent $-$ To interpret the extracted absorption and emission line
profiles, we propose that the outflow from \obj\ is constituted of a
high-velocity polar flow (at the origin of the intrinsic variations)
and a dense disk expanding at lower velocity and seen nearly
edge-on. This is in agreement with spectropolarimetric data and
supports the idea that BAL outflows can have large covering factors.

In our analysis, we focused on the most robust results.  Several
interesting questions nevertheless remain opened, requiring more
observations or a full radiative transfer modeling.  In particular,
why does the polarization of the different components apparently
differ?  Is the scattering region actually magnified?  High-quality
spectropolarimetry of the four images of \obj\ may solve this issue
and bring more informations on the nature of the ouflow.  As seen in
Figs. \ref{fig:sepavis} to \ref{fig:sepac}, different parts of the
emission line profiles appear micro-(de)amplified in components C and
D. Detailed line profile calculations are needed to understand the
origin of this micro-amplified emission and shed light on the outflow
kinematics.  With excellent signal-to-noise spectra, it would also be
possible to better use the microlensing effect observed in component
C.

\begin{acknowledgements}
It is a pleasure to thank Virginie Chantry for providing us with the
image illustrated in Fig.~\ref{fig:image}.  A fellowship from the
Alexander von Humboldt Foundation to DS is gratefully acknowledged.
\end{acknowledgements}

\appendix

\section{Example of a line profile decomposition}
\label{app:a1}

We consider a line profile constituted of an underlying continuum
$F_c(\lambda)$ (absorbed or not at some wavelengths) and an emission
profile $F_e(\lambda) = E_a(\lambda) + E_b(\lambda)$. We assume the
continuum $F_c$ micro-amplified by a constant factor $\mu_c$, the
component $E_a$ of the emission micro-amplified by a constant factor
$\mu_e$, and the component $E_b$ unaffected by
microlensing\footnote{It is always possible to write
$\mu'_e(\lambda)\; F_e = \mu_e \, E_a + E_b$ where $\mu_e$ =
max($\mu'_e(\lambda)$).}.  We consider a typical case with $\mu_c >
\mu_e > 1$.  If $M$ is the relative macro-amplification factor between
images 1 and 2, we have
\begin{eqnarray} 
F_1 & = & M \mu_c \, F_c + M \mu_e \, E_a + M \, E_b 
\\ 
F_2 & = &   F_c + E_a + E_b     \; . 
\end{eqnarray} 
Using these expressions with $A = M \mu_c$ and $\mu = \mu_c$ in Eqs.~4
and~5, we find
\begin{eqnarray} 
F_M \ & = & E_b + \frac{\mu_c - \mu_e}{\mu_c-1} \, E_a
\\ 
F_{M\mu}
& = &  F_c + \frac{\mu_e - 1}{\mu_c-1} \, E_a  \; .
\end{eqnarray} 
With $\mu_c > \mu_e > 1$, $ F_M > 0$ and $F_{M\mu}> 0$.  As expected,
$F_M$ contains the emission profile $E_b$ unaffected by microlensing,
$F_{M\mu}$ contains the continuum, and both $F_M$ and $F_{M\mu}$
contain a part of the micro-amplified emission profile $E_a$.  Up to a
scaling factor, the micro-amplified profile $E_a$ is given by
$F_{M\mu} - F_c$.

To effectively compute Eqs.~4 and 5 and to determine the
profile of $F_{M\mu}$, we need to know $M\,$\footnote{As
discussed in Sect.~\ref{sec:method}, the factor $A$ and then the
profile of $F_{M}$ are more easily determined.} (or $\mu_c$).
Practically, we consider $M$ as a free parameter in
Eq.~5 and, by varying it, we adopt the value of $M$ closest to~$A$
such that $F_{M\mu}(M) \geq F_c$ at all wavelengths (Eq.~7).  This is
equivalent to adopt a value of $\mu_c$ as close as possible to 1, thus
ensuring that the macro- and micro-amplifications are best separated
(see also Sluse et al.~\cite{slu07}).  Unless $M \simeq A$, this
method provides a reasonably accurate estimate of $M$, and then of
$\mu_c$.  Indeed, denoting the free parameter $M'$ and using
expressions A.1 and A.2 in Eq.~5, we write
\begin{eqnarray} 
F_{M\mu}(M') -F_c
& = &  \frac{M \mu_e - M'}{A-M} \, E_a + \frac{M - M'}{A-M} \, E_b   
\end{eqnarray} 
such that $F_{M\mu}(M') \geq F_c$ is satisfied when
\begin{eqnarray} 
M' &  = & M \, \mu_{\rm min}
\end{eqnarray} 
with
\begin{eqnarray} 
\mu_{\rm min} &  = & \min \left(\frac{\mu_e E_a+E_b}{E_a + E_b}\right) \ \ \ \ \ \  {\rm over\ the\  line\ profile.}
\end{eqnarray} 
If $E_a=0$ and $E_b \ne 0$ at some wavelengths, then $\mu_{\rm min}
=1$ and $M = M'$.  This means that $M$ can be derived empirically when
at least a portion of the observed emission line profile $F_e$ is not
micro-amplified.  In fact this is formally always true.  Indeed, if
$\mu_{\rm min} \neq 1$, the whole profile would be micro-amplified by
$\mu_{\rm min}$ i.e.  macro-amplified by $M\mu_{\rm min}$ instead of
$M$, which is inconsistent with an optimal separation of macro- and
micro-lensing\footnote{The amplification factors can be renormalized
as follows: $M' = M \mu_{\rm min}$, $\mu'_c = \mu_c / \mu_{\rm min}$
and $\mu''_e(\lambda) = \mu'_e(\lambda) / \mu_{\rm min}$. Note that
$\mu_{\rm min}$ = min($\mu'_e(\lambda))$.}. It should nevertheless be
kept in mind that if the emission line is amplified just like the
continuum, microlensing cannot be distinguished from macrolensing.
Fortunately, it is realistic to assume that a portion --even a very
small one-- of the observed emission line profile $F_e$ is not
micro-amplified given the large size of the emission line region
compared to the source of the continuum and to a typical Einstein
radius.  This hypothesis can be verified as follows: if we notice that
$F_{M\mu} = F_c$ and $F_M \ne 0$ over a certain wavelength range, then
we must have $M'$ = $M$ and $\mu_{\rm min} = 1$. Indeed, combining
Eqs. A.5 and A.6, we find that when $F_{M\mu} = F_c$ then either $E_a$
must be proportional to $E_b$ over that wavelength range --which is
unlikely since $E_a$ and $E_b$ are meant to originate from different
emission regions-- or $\mu_{\rm min} = 1$.

\section{The absorption / emission toy model}
\label{app:a2}

For a given line profile, we adopt for the disk optical depth $\tau_d$ the
functional form
\begin{eqnarray}
\tau_d(w) = \sum_{i} \tau_{0}(i) \; \exp -| \frac{w-w_{ca}(i)}{w_{sa}} |
\end{eqnarray}
where $w$ is the velocity. $w_{ca} (i)$ is the position of the center
of the absorption line $i$, and $w_{sa}$ the width taken identical for
all lines.  The sum is computed over the 2 or 3 lines which constitute
a line profile, with the appropriate weights $\tau_{0}(i)$.
Similarly, we adopt for the unabsorbed emission line profile
\begin{eqnarray}
f_{e}(w) =  \sum_{i} f_{e0}(i) \; \exp - \left( \frac{w-w_{ce}(i)}{w_{se}} 
\right) ^2 \; .
\end{eqnarray}
The disk absorption profile is computed as $F_a = \exp(-\tau_d)$ and
the emission line profile absorbed by the disk as $F_{ea} =
f_{e}\,\exp(-\tau_d)$. The total absorption of the continuum must be
equal to the absorption in the polar flow multiplied by the absorption
in the disk.  Although not computed, it can be obtained by inverting
the observed intrinsic absorption profile.

For all line profiles shown in Fig.~\ref{fig:vel}, we use $w_{ca}(1)$
= $-$5000 km s$^{-1}$, $w_{sa}$ = 800 km s$^{-1}$, $w_{ce}(1)$ =
$-$2000 km s$^{-1}$, $w_{se}$ = 5000 km s$^{-1}$, $i=1$ corresponding
to the reddest line of the profiles, the position of the other ones
being fixed by the doublet separation and/or by the Ly$\alpha$ --
\ion{N}{v} velocity separation. $\tau_{0}(i)$ and $f_{e0}(i)$ are
choosen to fit the observations.  The parameters are not unique and
may be different at other epochs.


\begin{thebibliography}{999}

\bibitem[1990]{ang90} Angonin, M.-C., Vanderriest, C., Remy, M., 
Surdej, J.\ 1990, \aap, 233, L5

\bibitem[2008]{ang08} Anguita, T., Faure, C., Yonehara, A.,
Wambsganss, J., Kneib, J.-P., Covone, G.,  Alloin, D.\ 2008, \aap,
481, 615

\bibitem[2008b]{ang08b} Anguita, T., Schmidt, R.~W., Turner, E.~L.,
Wambsganss, J., Webster, R.~L., Loomis, K.~A., Long, D.,  McMillan,
R.\ 2008, \aap, 480, 327

\bibitem[1989]{bar89} Barlow, T.~A., Junkkarinen, V.~T.,  Burbidge,
E.~M.\ 1989, \apj, 347, 674

\bibitem[1992]{bar92} Barlow, T.~A., Junkkarinen, V.~T., Burbidge,
E.~M., Weymann, R.~J., Morris, S.~L.,  Korista, K.~T.\ 1992, \apj,
397, 81

\bibitem[2000]{bec00} Becker, R.~H., White, R.~L., Gregg, M.~D.,
Brotherton, M.~S., Laurent-Muehleisen, S.~A.,  Arav, N.\ 2000, \apj,
538, 72

\bibitem[2000]{bel00} Belle, K.~E.,  Lewis, G.~F.\ 2000, \pasp, 112,
320

\bibitem[1994]{bjo94} Bjorkman, J.~E., Ignace, R., Tripp, T.~M., 
Cassinelli, J.~P.\ 1994, \apj, 435, 416

\bibitem[2010]{bor10} Borguet, B., Hutsem{\'e}kers, D. \ 2010, \aap, 515, A22

\bibitem[2008]{bor08} Borguet, B., Hutsem{\'e}kers, D., Letawe, G.,
Letawe, Y., Magain, P.\ 2008, \aap, 478, 321

\bibitem[2009]{bra09} Bradford, C.~M., et al.\ 2009, arXiv:0908.1818

\bibitem[2001]{bur01} Burud, I.\ 2001, Ph.D.~Thesis,  

\bibitem[1999]{cha99} Chae, K.-H., Turnshek, D.~A.\ 1999, \apj, 514,
587

\bibitem[2001]{cha01} Chae, K.-H., Turnshek, D.~A., Schulte-Ladbeck,
R.~E., Rao, S.~M.,  Lupie, O.~L.\ 2001, \apj, 561, 653

\bibitem[2007]{cha07} Chantry, V., Magain, P.\ 2007, \aap, 470, 467

\bibitem[2005]{che05} Chelouche, D.\ 2005, \apj, 629, 667

\bibitem[1990]{cor90} Corbin, M.~R.\ 1990, \apj, 357, 346

\bibitem[2008]{eig08} Eigenbrod, A., Courbin, F., Meylan, G., Agol,
E., Anguita, T., Schmidt, R.~W., \& Wambsganss, J.\ 2008, \aap, 490,
933

\bibitem[2001]{fur01} Furlanetto, S.~R.,  Loeb, A.\ 2001, \apj, 556,
619

\bibitem[2007]{gal07} Gallagher, S. C., Hines, D. C., Blaylock, Myra, 
Priddey, R. S., Brandt, W. N., Egami, E. E. \ 2007, \apj, 556,
619

\bibitem[2008]{gib08} Gibson, R.~R., Brandt, W.~N., Schneider, D.~P.,
 Gallagher, S.~C.\ 2008, \apj, 675, 985

\bibitem[2010]{goi10} Goicoechea, L.J., Shalyapin, V.N. \ 2010, 
\apj, 708, 995 

\bibitem[1997]{goo97} Goodrich, R.~W.\ 1997, \apj, 
474, 606 

\bibitem[1995]{goo95} Goodrich, R.~W.,  Miller, J.~S.\ 1995, \apjl,
448, L73

\bibitem[1993]{ham93} Hamann, F., Korista, K.~T.,  Morris, S.~L.\
1993, \apj, 415, 541

\bibitem[1990]{hut90} Hutsem\'ekers, D., Surdej, J. 1990, \apj , 
361, 367

\bibitem[1993]{hut93} Hutsem\'ekers, D. 1993, \aap , 280, 435

\bibitem[1998]{hut98} Hutsem\'ekers, D., Lamy, H.,  Remy, M.\ 1998,
\aap, 340, 371

\bibitem[1994]{hut94} Hutsem\'ekers, D., Surdej, J.,  van Drom, E.\
1994, \apss, 216, 361

\bibitem[1990]{kay90} Kayser, R., Surdej, J., Condon, J.~J.,
Kellermann, K.~I., Magain, P., Remy, M.,  Smette, A.\ 1990, \apj,
364, 15

\bibitem[1998]{kne98} Kneib, J.-P., Alloin, D.,  Pello, R.\ 1998,
\aap, 339, L65

\bibitem[2002]{kur02} Kuraszkiewicz, J.~K., Green, P.~J., Forster, K.,
Aldcroft, T.~L., Evans, I.~N.,  Koratkar, A.\ 2002, \apjs, 143, 257

\bibitem[1999]{lam99} Lamers, H.J.G.L.M., Cassinelli, J.P. \ 1999,
Introduction to stellar winds, Cambridge University Press

\bibitem[2004]{lam04} Lamy, H., Hutsem{\'e}kers, D.\ 2004, \aap, 427,
107

\bibitem[1998]{lew98} Lewis, G.~F.,  Belle, K.~E.\ 1998, \mnras,
297, 69

\bibitem[2004]{lew04} Lewis, G.~F.,  Ibata, R.A. \ 2004, \mnras,
348, 24

\bibitem[2007]{lut07} Lutz, D., et al.\ 2007, \apjl, 661, L25

\bibitem[2009]{mac09} MacLeod, C.~L., Kochanek, C.~S.,  Agol, E.\
2009, \apj, 699, 1578

\bibitem[1988]{mag88} Magain, P., Surdej, J., Swings, J.-P., Borgeest,
U.,  Kayser, R.\ 1988, \nat, 334, 325

\bibitem[1998]{mag98} Magain, P., Courbin, F., Sohy, S.\ 1998, \apj,
494, 472

\bibitem[2009]{mar09} Markwardt, C.~B.\ 2009, arXiv:0902.2850

\bibitem[1999]{mci99} McIntosh, D.~H., Rix, H.-W., Rieke, M.~J.,
Foltz, C.~B.\ 1999, \apjl, 517, L73

\bibitem[1998]{mon98} Monier, E.~M., Turnshek, D.~A.,  Lupie, O.~L.\
1998, \apj, 496, 177

\bibitem[1995]{mur95} Murray, N., Chiang, J., Grossman, S.~A., 
Voit, G.~M.\ 1995, \apj, 451, 498

\bibitem[2008]{nes08} Nestor, D., Hamann, F., Rodriguez Hidalgo, P. 
\ 2008, \mnras, 386, 2055

\bibitem[1999]{ogl99} Ogle, P.~M., Cohen, M.~H., Miller, J.~S., Tran,
H.~D., Goodrich, R.~W.,  Martel, A.~R.\ 1999, \apjs, 125, 1

\bibitem[1997]{ost97} {\O}stensen, R., et al.\ 1997, \aaps, 126, 393

\bibitem[1992]{pei92} Pei, Y.~C.\ 1992, \apj, 395, 130

\bibitem[2008]{poi08} Poindexter, S., Morgan, N.,  Kochanek, C.~S.\
2008, \apj, 673, 34

\bibitem[2004]{pro04} Proga, D., Kallman, T.~R.\ 2004, \apj, 616, 688 

\bibitem[2003]{rei03} Reichard, T.~A., et al.\ 2003, \aj, 126, 2594

\bibitem[1996]{rem96} Remy, M., Gosset, E., Hutsem\'ekers, D.,
Revenaz, B.,  Surdej, J.\ 1996, Astrophysical Applications of
Gravitational Lensing, 173, 261

\bibitem[2002]{ric02} Richards, G.~T., Vanden Berk, D.~E., Reichard,
T.~A., Hall, P.~B., Schneider, D.~P., SubbaRao, M., Thakar, A.~R.,
York, D.~G.\ 2002, \aj, 124, 1

\bibitem[2004]{sah04} Saha, P., Williams, L.~L.~R.\ 2004, \aj, 127,
2604

\bibitem[2004]{sca04} Scannapieco, E.,  Oh, S.~P.\ 2004, \apj, 608,
62

\bibitem[2005]{sca05} Scannapieco, E., Silk, J., Bouwens, R.\ 2005,
\apjl, 635, L13

\bibitem[1999]{sch99} Schmidt, G.~D.,  Hines, D.~C.\ 1999, \apj,
512, 125

\bibitem[1992]{sch92} Schneider, P., Ehlers, J.,  Falco, E.~E.\
1992, Gravitational Lenses, Springer-Verlag Berlin Heidelberg New
York.

\bibitem[2008]{sch08} Schramm, M., Wisotzki,  L., Jahnke, K. \ 2008, 
\aap, 478, 311

\bibitem[1973]{sha73} Shakura, N.~I., Sunyaev, R.~A.\ 1973, \aap,
24, 337

\bibitem[1998]{sil98} Silk, J., Rees, M. \ 1993, \aap, 331, L1

\bibitem[2005]{slu05} Sluse, D., Hutsem{\'e}kers, D., Lamy, H.,
Cabanac, R., Quintana, H.\ 2005, \aap, 433, 757

\bibitem[2007]{slu07} Sluse, D., Claeskens, J.-F., Hutsem\'ekers, D.,
 Surdej, J.\ 2007, \aap, 468, 885

\bibitem[1987]{sur87} Surdej, J., Hutsem\'ekers, D.\ 1987, \aap, 177,
42

\bibitem[1988]{tur88} Turnshek, D.~A., Grillmair, C.~J., Foltz, C.~B.,
 Weymann, R.~J.\ 1988, \apj, 325, 651

\bibitem[1997]{tur97} Turnshek, D.~A., Lupie, O.~L., Rao, S.~M.,
Espey, B.~R.,  Sirola, C.~J.\ 1997, \apj, 485, 100

\bibitem[1993]{voi93} Voit, G.~M., Weymann, R.~J.,  Korista, K.~T.\
1993, \apj, 413, 95

\bibitem[1993]{wit93} Witt, H.~J., Kayser, R., Refsdal, S. \ 1993,
\aap, 268, 501

\bibitem[1991]{wey91} Weymann, R.~J., Morris, 
S.~L., Foltz, C.~B.,  Hewett, P.~C.\ 1991, \apj, 373, 23 

\bibitem[2007]{you07} Young, S., Axon, D.~J., Robinson, A., Hough,
J.~H.,  Smith, J.~E.\ 2007, \nat, 450, 74

\bibitem[2006]{zho06} Zhou, H., Wang, T., Wang, H., Wang, J., Yuan,
W.,  Lu, Y.\ 2006, \apj, 639, 716

\end{thebibliography}
\end{document}